\documentclass[12pt,a4paper]{article}

\usepackage{amsmath,amssymb,amsfonts}
\usepackage[dvips]{graphicx}
\usepackage{epsfig}
\usepackage{pdfsync}
\usepackage{cite}
\usepackage{bbm}
\usepackage{braket}
\usepackage{eufrak}
\usepackage{multirow}
\usepackage{verbatim}
\usepackage{hyperref}

\usepackage[left=2.8cm,top=3.3cm,right=2.8cm,bottom=3.3cm,bindingoffset=0cm]{geometry}
\newcommand{\p}{\partial}
\newcommand{\be}{\begin{equation}}
\newcommand{\ee}{\end{equation}}
\newcommand{\D}{\mathcal{D}}
\newcommand{\beq}{\begin{eqnarray}} 
\newcommand{\eeq}{\end{eqnarray}}
\def\de{\partial}

\DeclareMathOperator{\Tr}{Tr}
\DeclareMathAlphabet{\mathpzc}{OT1}{pzc}{m}{it}

\renewcommand\Re{\operatorname{Re}}
\renewcommand\Im{\operatorname{Im}}

\def\Tr{ \hbox{\rm Tr}}

\def\de{\partial}

\def\half{\frac{1}{2}}

\makeatother

\def\nn{\nonumber\\}
\newdimen\Tdim
\def\ispan{{\setbox0=\hbox{i}%
\Tdim\ht0\advance\Tdim\dp0\rule[-\dp0]{0pt}{\Tdim}}}
\def\jspan{{\setbox0=\hbox{j}%
\Tdim\ht0\advance\Tdim\dp0\rule[-\dp0]{0pt}{\Tdim}}}
\def\Tspan#1{{\setbox0=\hbox{#1}%
\Tdim\ht0\advance\Tdim\dp0\advance\Tdim.55ex\rule[-\dp0]{0pt}{\Tdim}\box0}}

\numberwithin{equation}{section}

\begin{document}

\title{
\begin{flushright}\ \vskip -1.5cm {\small{IFUP-TH-2015}}\end{flushright}
\vskip 18pt
\bf{ \Large  Geometry and  Dynamics of  a   \\
Coupled $4D$-$2D$  Quantum Field Theory
 }
\vskip 14pt}
\author{
Stefano Bolognesi$^{(1,2)}$, Chandrasekhar  Chatterjee$^{(2,1)}$,  Jarah Evslin$^{(3,2)}$, \\
Kenichi Konishi$^{(1,2)}$, Keisuke Ohashi$^{(1,2,4)}$,  Luigi Seveso$^{(1)}$  \\[20pt]
{\em \small
$^{(1)}$Department of Physics ``E. Fermi", University of Pisa}\\[0pt]
{\em \small
Largo Pontecorvo, 3, Ed. C, 56127 Pisa, Italy}\\[3pt]
{\em \small
$^{(2)}$INFN, Sezione di Pisa,    
Largo Pontecorvo, 3, Ed. C, 56127 Pisa, Italy}\\[3pt]
{\em \small
$^{(3)}$ NanChangLu 509, Inst. Mod. Phys., CAS,  
730000 Lanzhou, China}   \\[3pt]
{\em \small
$^{(4)}$ Osaka City University Advanced Mathematical  Institute, } \\ [3pt] 
{\em \small 
3-3-138 Sugimoto, Sumiyoshi-ku, Osaka 558-8585, Japan} \\ [3pt] 
{ \footnotesize Emails: stefanobolo@gmail.com, chandra.chttrj@gmail.com,}\\
{ \footnotesize  jarah@impcas.ac.cn, kenichi.konishi@unipi.it,}   \\
{ \footnotesize   keisuke084@gmail.com, keisuke.ohashi@for.unipi.it, sevesoluigi@gmail.com}
}
\vskip 4pt
\date{September 2015}
\maketitle

\begin{center}

{\large Abstract}

\end{center}

Geometric  and dynamical aspects of a coupled $4D-2D$ interacting quantum field theory - the gauged nonAbelian vortex -  are investigated. The fluctuations of the internal $2D$ nonAbelian vortex zeromodes excite the massless $4D$ Yang-Mills modes and  in general give rise to divergent energies. 
This  means that the well-known $2D$ $\mathbbm {CP}^{N-1}$ zeromodes associated with a nonAbelian vortex become nonnormalizable.
%To compensate the freezeout of $2D$ quantum fluctuations, 
Moreover, all sorts of global, topological  $4D$  effects such as
the nonAbelian Aharonov-Bohm effect come into play.  These 
topological global features and the 
dynamical properties associated with the fluctuation of the $2D$ vortex moduli modes are intimately correlated, as shown
concretely here in a $U_0(1)\times SU_l(N)\times SU_r(N)$  model with scalar fields in a bifundamental representation of the two $SU(N)$ factor gauge groups.

\newpage

\section{Introduction}

NonAbelian vortices - vortex solutions carrying nonAbelian continuous orientational zeromodes  $\{ B_{i} \}$  -  have been extensively investigated in the last decade, revealing many interesting features arising from the soliton and gauge dynamics, topology,  and 
 global symmetries \cite{Auzzi:2003fs}-\cite{Gudnason:2010rm}.  Typically they occur in a system in the color-flavor locked phase, i.e., systems in which the gauge symmetry is broken by a set of scalar condensates that, however, leave a color-flavor diagonal symmetry intact.
Color-flavor locked systems appear to be quite ubiquitous in Nature. Standard QCD at zero temperature exhibits some characteristic features of this sort, as  can be seen in a hidden-symmetry perspective \cite{Komargo}.   
They occur in the infrared effective theories of many ${\cal N}=2$ supersymmetric 
theories softly broken to ${\cal N}=1$, and may carry important hints about  the mechanism responsible for quark confinement \cite{CKM}. In particular they could 
shed light on the mysteries of nonAbelian monopoles. They are realized in high-density QCD in the color superconductor phase \cite{ColorSC}, which may well be realized in the interiors of neutron stars. 
%Related phenomena could also find interesting applications in the context of Bose-Einstein condensed cold atom systems or in some multi-component superconductors. 

Other fascinating aspects associated with these vortex solutions appear  when further gauge fields  are introduced, coupled to  part or all of the color-flavor diagonal global symmetry.  We shall refer to these systems as ``gauged nonAbelian vortices'' in this paper.  All sorts of global effects, such as nonAbelian Aharonov-Bohm phases and scattering,  an obstruction to part of the ``unbroken" gauge symmetry,   nonAbelian statistics under the exchange of parallel vortices, Cheshire charges, etc. make their appearance, depending on the vortex orientations, $\{B_i \}$.  
These phenomena have been investigated in various general contexts \cite{Alford:1990mk}-\cite{Bucher:1993jj}    with gauge symmetry breaking, $G \to H$, with  $\pi_1(G/H)\ne  {\bf 1}$, and more recently, in the context of concrete model, e.g., in a $U_0(1)\times SU_l(N)
\times SU_r(N)$ gauge theory, with scalar fields in the bifundamental representation of the two $SU(N)$ gauge groups \cite{Konishi:2012eq}-\cite{Bolognesi:2014saa}
.     

A sort of paradox or dilemma  seems to arise, however.   One of  the most important, characteristic  features of nonAbelian vortices 
is their collective  dynamics. Quantum fluctuations of the vortex internal orientational modes are described by various $2D$  (vortex worldsheet) sigma models, such as  ${\mathbbm {CP}}^{N-1}$ \footnote{This occurs in a model  with $U_0(1)\times SU(N)$ gauge theory.   Similar models, involving color-flavor locked vacua with $SO(2N)$ or $USp(2N)$ gauge symmetry, yield sigma models with target  Hermitian symmetric spaces,  such as $SO(2N)/U(N)$,  $USp(2N)/U(N)$,  etc. }.    The ${\mathbbm {CP}}^{N-1}$ interactions  are asymptotically free: 
the vortex orientation $\{B_i \}$ fluctuates strongly at large distances.  As a result, all of the global topological  effects mentioned above would be washed away.

Let us remind ourselves that  a characteristic feature of a color-flavor locked vacuum  is the fact that all massless  Nambu-Goldstone particles are eaten by the broken gauge fields, all of which become massive, maintaining mass degeneracy among them. No massless scalars or gauge bosons survive in the bulk.  In the vortex sector, the only massless modes are those of the vortex orientational zeromodes, which are a kind of Nambu-Goldstone excitation, confined within the vortex worldsheet.
 In the case of gauged nonAbelian vortices, instead,  some combinations of the gauge bosons remain massless in the $4D$ bulk, and their coupling to the fluctuations of the $2D$ zeromodes $\{ B_{i} \}$  are expected to affect significantly the dynamics of the  latter.    Some preliminary studies of these issues have been done  \cite{Konishi:2012eq,Evslin:2013wka}.

  It is the purpose of the present article to examine more thoroughly the effects of the unbroken $4D$ gauge interactions on the gauged nonAbelian vortex collective dynamics  in the $2D$ vortex worldsheet.  A nontrivial task is that of disentangling the effects of the extra gauge fields on the static vortex configuration itself from those of  residual  dynamical effects of the  $4D$ massless gauge modes and their  couplings to  the massless $2D$ orientational modes. 
These problems will be worked out systematically below. 
 
This paper is organized as follows.  In Section~\ref{model} the model  and the main properties of the vortex solution are presented together with a  brief review of the associated  global effects.
% are mentioned in Section~\ref{sec:degeneracy}.   
The closely related question of topological and geometric obstructions is reviewed, generally and in our concrete model,  in Section~\ref{sec:topological}. 
Vortex fluctuations and the induced excitations of the $4D$  Yang-Mills modes are worked out in Section~\ref{sec:excitation} and  Section~\ref{sec:energy}. Section~\ref{sec:origin} is dedicated to clarifying the connection between the global, topological aspects of the gauge-vortex  system and the {\it dynamics} of the zeromode excitations of the vortex. Finally, in Section~\ref{sec:discussion} we examine the apparent discontinuity in physics in the limit in which one of the nonAbelian gauge factors,  $SU_r(N)$ (or $SU_l(N)$),  is decoupled. It is argued that it is essentially due to the noncommutativity of the two limits, $g_r\to 0$ and  $R \to \infty$, where $R$ is an infrared cutoff introduced to regularize the energy divergences caused by the vortex fluctuations. 
Appendix A deals with the peculiarity of the solution of Gauss's equations in the particular $g_r=0$ case. Appendix B   proves the uniqueness of the Ansatz Eq.~(\ref{color1}) used to solve Gauss's equations in Section~\ref{sec:excitation} and  Section~\ref{sec:energy}.

\section{The model, vortex solutions and AB effect \label{model}}

Even though our study can be generalized to the case of an arbitrary gauge group of the type
\[   G=   U_{0}(1) \times  G_{L}\times G_{R}\ ,   
\]
 we shall choose,  for concreteness, to work with   $G_{L}=G_{R}= SU(N)$.  The matter sector consists of a complex scalar field  $Q$
in the bifundamental representation of the two $SU(N)$ factors, with unit charge with respect to $U_{0}(1)$.

\subsection{Vortex solutions}

We shall work with 
a BPS-saturated action \footnote{This is, after adding the appropriate adjoint scalar fields (not relevant for the vortex solution hence set to zero),  the truncated bosonic sector of a ${\cal N}=2$ supersymmetric theory.}
 \beq
\label{action}
{\cal L} &=& -\frac{1}{2} \Tr \, (F^{(l)}_{\mu\nu}F^{(l)\mu\nu})  -\frac{1}{2} \Tr \, (F^{(r)}_{\mu\nu}F^{(r)\mu\nu}) -\frac{1}{4}  \, f_{\mu\nu} f^{\mu\nu}  +  \Tr\,(D_{\mu} Q^{\dagger}D^\mu Q)   \nonumber \\
 &&  -  \frac{g_0^2}{2}    ( \Tr\,{ Q}^{\dagger} Q    -  v_0^2)^2    -  \frac{g_{l}^2}  {2}     ( \Tr\,t^a  QQ^{\dagger} )^2   - \frac{g_r^2}{2}    ( \Tr\, t^a  Q^\dag Q)^2\ , 
\eeq
where $v_0^2= N \xi$ and the covariant derivative is
%\footnote{The sign in front of $g_r$ is purely conventional.}
\be
D_{\mu} Q = \partial_{\mu} Q - i  g_{l}  A^{(l)}_{\mu} Q - i  g_{0} a_{\mu} Q   + i  g_r Q  A^{(r)}_{\mu} \ . \label{dq}
\ee
The scalar-field condensate in the vacuum takes the form
\beq
\langle Q \rangle =\sqrt{\xi} \, {\bf 1}_{N}\ ,
\label{qvacuum}
\eeq
leaving a left-right diagonal $SU(N)$ gauge group unbroken.
The fields 
\beq
\label{masslesscombination}
{\cal A}_{\mu}=  \frac{1}{\sqrt{g_{r}^2 + g_{l}^2}  }\,( g_{r}   A^{(l)}_{\mu}   +     g_{l}  A^{(r)}_{\mu} )\ ,\eeq 
remain massless in the bulk, 
whereas  the orthogonal combination
\beq
\label{massivecombination}
{\cal B}_{\mu} =  \frac{1}{\sqrt{g_{r}^2 + g_{l}^2}  }\,( g_{l}   A^{(l)}_{\mu}   -    g_{r}  A^{(r)}_{\mu} ) 
\eeq 
and the $U(1)$ field $a_{\mu}$, become massive.

The nontrivial  first homotopy group
\be   \pi_1 \left(   \frac{ U_{0}(1) \times SU_L(N)\times SU_{R}(N) }{ SU_{L+R}(N)} \right) = {\mathbbm {Z}} \ , 
\ee
means that the system admits stable vortices, which are our main interest below. 
The vortex solutions can be found by the BPS completion of the expression for the tension
(for configurations depending only on  the transverse coordinates $x$ and $y$):
\begin{eqnarray}
T   & = &   \int d^{2}x\,  \Bigg\{\,  \frac{1}{2} \left( f_{12}  + {g_{0}}  (\Tr \, Q^{\dagger}   Q -  N \xi)\right)^{2} +   \Tr \left[\left( F^{(r)}_{12}  -  g_{r}  \,t^a  \Tr\,(Q^{\dagger} Q  t^a)  \right)^{2}+\right. \nonumber \\
 &&   \ \ \  + \left.     \left( F^{(l) }_{12} +  g_{l} \, t^a  \Tr\,( Q^{\dagger}  \,  t^a   Q )  \right)^{2}  \right]   +      \Tr\,|D_{1} Q +  i D_{2} Q|^{2} +   {g_{0} N }   \, \xi\,  f_{12}\Bigg\}\ .
%  -  \left.\frac12\epsilon_{ij}\partial_{i}\left(i \nabla_{j}Q\bar Q- Q\,i\nabla_{j}\bar Q \right) \right\}
 \label{BPScompletion} 
\end{eqnarray}
The BPS equations are accordingly:
\begin{eqnarray}
&&   D_{1} Q  +  i D_{2} Q=0\ , \qquad\qquad\qquad
  f_{12}  +    {g_{0}}  (\Tr \, Q^{\dagger}   Q -  N \xi) =0\ ,   \label{BPSequations2} \\
 &&          F^{(r)}_{12}  - g_{r}  \,t^a  \Tr( Q^{\dagger} Q  t^a)      =0\ ,\qquad
          F^{(l) }_{12} + g_{l} \, t^a  \Tr( Q^{\dagger}  \,  t^a   Q )     =0\ .   \label{BPSequations4}
\end{eqnarray}
For a minimal vortex with a fixed orientation in color-flavor, for example $\left(1,{\bf 1}_{N-1}\right)$, one can take 
the scalar field Ansatz to be \footnote{We consider the minimal winding vortex below;  the generalization
of the formulas below to higher-winding solutions is straightforward.}
\beq
Q=   \left(\begin{array}{cc}e^{i \theta} \, Q_1(r) & 0 \\0 & Q_2(r)\, {\bf 1}_{N-1}\end{array}\right)\ ,    
 \label{vortex}
\eeq
whereas the nonAbelian and Abelian gauge fields can be written in the diagonal form 
\be
a_i=-\frac{1}{g_0}\frac{\epsilon_{ij}x_j}{r^2}\frac{1-f}{N}\,;
\label{A0}  \ee
\be
A_i^{(l)}=-\frac{g_l}{g'^2}\frac{\epsilon_{ij}x_j}{r^2}\frac{1-f_{NA}}{N C_N} T_{N^2-1}\,;   \label{Al}
\ee
\be
A_i^{(r)}=\frac{g_r}{g'^2}\frac{\epsilon_{ij}x_j}{r^2}\frac{1-f_{NA}}{N C_N} T_{N^2-1}\ ,\label{Ar}
\ee
where
\beq     T_{N^2-1} \equiv   C_N  \,  \left(\begin{array}{cc}N-1 & 0 \\0 & -{\bf 1}_{N-1}\end{array}\right)\ , \qquad   C_N\equiv \frac{1}{\sqrt{2N(N-1)} }\ ,
\eeq
and 
\be   g^{\prime} \equiv \sqrt{g_{l}^{2}+ g_{r}^{2}}\ .  
\ee
The boundary conditions are 
\be    f(0) =  f_{NA}(0)=1 \ , \qquad f(\infty)=  f_{NA}(\infty)=0\ , \nonumber \ee
\be   Q_1(\infty)=  Q_2(\infty) =  \sqrt{\xi} \ .  \label{boundary}\ee
The BPS equations (\ref{BPSequations2})-(\ref{BPSequations4}) then show that the profile functions satisfy 
\be
\label{bpsvortexuno}
\frac{f'}{r}-g_0^2N[Q_1^2+(N-1)Q_2^2-   N \xi ]=0\ ,
\ee
\be 
\label{bpsvortexdue}
\frac{f'_{NA}}{r}-g'^2\frac{Q_1^2-Q_2^2}{2}=0\ ,
\ee
\be 
\label{bpsvortextre}
rQ_1'-Q_1\left(\frac{(N-1)f_{NA}+f}{N}\right)=0\ , 
\qquad 
rQ_2'-Q_2\left(\frac{-f_{NA}+f}{N}\right)=0\ ,  
\ee
which can be solved  by  numerical methods.  
These equations are identical to those found earlier for the global nonAbelian vortex, i.e., for $g_{r}=0$ or $g_{l}=0$, except for the fact that  
the gauge fields compensating the scalar winding  energy $\de Q / \de \theta$ are now shared between  the left and right $SU(N)$ fields.  
In other words, the static vortex profile remains basically 
unmodified as compared to the standard nonAbelian vortex (and for that matter, the ANO vortex), with a well-defined width, $\sim \tfrac{1}{\sqrt{\xi}}$. 
The vortex tension is given by (see Eq.~(\ref{BPScompletion})) 
\be   T = 2\pi \xi\ . \label{tension}
\ee 

\subsection{The Aharonov-Bohm effect}

Whenever the (untraced) Wilson loop in some representation of a gauge field around the vortex is not equal to unity, particles belonging to that representation are transformed when encircling the vortex.  The transformation is given by the (untraced) Wilson loop.  This is called the Aharonov-Bohm (AB) effect.

These Wilson loops are easily calculated from Eqs.~(\ref{A0}-\ref{Ar})
\be
\lim_{r\rightarrow\infty}  {\rm{exp}}\left({i g_0\oint a}\right)={\rm{exp}}\left({\frac{2\pi i}{N}{\bf 1}_{N}}\right)\ , \label{loop0}
\ee
\be
\lim_{r\rightarrow\infty}{\rm{exp}}\left({i g_l \oint A^{(l)}}\right)={\rm{exp}}\left({2\pi i\frac{g_l^2}{g^{\prime 2}}} \left(\begin{array}{cc}1-\frac{1}{N} & 0 \\0 & -\frac{1}{N}{\bf 1}_{N-1}\end{array}\right)\right)\,; \label{loopl}
\ee
\be
\lim_{r\rightarrow\infty}{\rm{exp}}\left({i g_r \oint A^{(r)}}\right)={\rm{exp}}\left({-2\pi i\frac{g_r^2}{g^{\prime 2}}} \left(\begin{array}{cc}1-\frac{1}{N} & 0 \\0 & -\frac{1}{N}{\bf 1}_{N-1}\end{array}\right)\right).  \label{loopr}
\ee
To obtain the AB phase given a representation of $G$ is now quite straightforward.  For example, consider the scalar field $Q$, which is of charge 1 under $U(1)_0$, transforms in the fundamental of $SU(N)_l$ and the antifundamental of $SU(N)_R$.  The corresponding AB phase is simply the product of (\ref{loop0}) and (\ref{loopl}) divided by (\ref{loopr}), which is the identity, so $Q$ does not transform.  This is an important consistency check because the $Q$ field condenses and any condensate must be single valued.  

The AB phase can be calculated using the gauge fields in the ultraviolet $A^{(l)}$ and $A^{(r)}$ basis or in the mass eigenstate ${\cal{A}}$ and ${\cal{B}}$ basis.   Of course the two answers must agree.  To use the second basis, one needs the corresponding Wilson loops, which are the exponentials of
\be
\lim_{r\rightarrow\infty}\,{i \frac{g_l g_r}{g^\prime} \oint {\cal{A}}}
=\lim_{r\rightarrow\infty}\,{ i  \frac{g_l g_r}{{g^\prime}^2} \oint\left(g_r A^{(l)}+g_lA^{(r)}\right)}
= {\bf 0}_N\ ,
\ee
\be
\lim_{r\rightarrow\infty}\,{i g^\prime\oint  {\cal{B}}}
=\lim_{r\rightarrow\infty}\,{i\oint  \left(g_l A^{(l)}-g_rA^{(r)}\right)}
=2\pi i \left(\begin{array}{cc}1-\frac{1}{N} & 0 \\0 & -\frac{1}{N}{\bf 1}_{N-1}\end{array}\right) .
\ee
The fact that $\oint\cal{A}$ vanishes implies that the AB phase is independent of the $\cal{A}$ charge, it only depends on the charge under the massive gauge field ${\cal{B}}$ and $U(1)_0$.  

Rewriting (\ref{dq}) at spatial infinity as
\be
D_{\mu} Q = \partial_{\mu} Q - i  g^\prime {\cal{B}}_\mu Q - i  g_{0} a_{\mu} Q  \ ,
\ee
one sees that in the mass basis $Q$ has charge 1 under $U(1)_0$ and also under ${\cal{B}}$.  Therefore upon circumnavigating a vortex at large radius, $Q$ is rotated by the exponential of 
\be
\lim_{r\rightarrow\infty}{\oint i (g_0 a+g^\prime {\cal{B}})}
%=-\frac{2\pi i}{N} {\bf 1}_N-2\pi i \left(\begin{array}{cc}1-\frac{1}{N} & 0 \\0 & -\frac{1}{N}{\bf 1}_{N-1}\end{array}\right) 
=2\pi i \left(\begin{array}{cc}-1 & 0 \\0 & {\bf 0}_{N-1}\end{array}\right). \label{quant}
\ee
This exponential is unity and so again we recover the fact that $Q$ is invariant.  The invariance of $Q$, which is necessary for the condensate to be well-defined, imposes that the matrix in Eq.~(\ref{quant}) is integral.  This in fact is the quantization condition on the nonAbelian vortex charge.  In the low energy gauge theory it is the topological condition that the flux of a Higgsed gauge field is quantized.  In particular, it is preserved by any continuous deformation of the vortex.

\subsection{Vortex zeromodes  (degeneracy) \label{sec:degeneracy}}

The solution (\ref{vortex}) further  breaks the unbroken color-flavor diagonal $H=  SU_{l+r}(N)$ group as 
\be      H=SU(N) \to {\widetilde H} = U(N-1)\ ,
\ee
implying the existence of degenerate solutions, corresponding to the coset, 
\be   H/ {\widetilde H} =  SU(N) / U(N-1)   \sim   {\mathbbm {CP}}^{N-1}\ .
\ee 
The general solutions are related to (\ref{vortex})  by global  $SU_{l+r}(N)$ transformations,  
\be   \{  Q,   A_i^{(l)}, A_i^{(r)} \} \to    U\,  \{  Q,   A_i^{(l)}, A_i^{(r)} \} \, U^{\dagger}\ ,  \label{zeromodes}
\ee
i.e.,  
\be      Q^{(B)}=    U(B) \,  \left(\begin{array}{cc}e^{i \theta} \, Q_1(r) & 0 \\0 & Q_2(r)\, {\bf 1}_{N-1}\end{array}\right)  \, U^{\dagger}(B)\ ,  
\ee  
\be    A_i^{(l,\, B)} =      U(B) \, A_i^{(l)} \, U^{\dagger}(B)\ ,\qquad  A_i^{(r,\, B)} =      U(B) \, A_i^{(r)} \, U^{\dagger}(B)\ , 
\ee
where the rotation matrix
\be  U = 
\left(
\begin{array}{cc}
 X^{-\half} & - B^\dagger Y^{-\half}  \\
 X^{-\half}B &  Y^{-\half}   
\end{array}
\right)\ , \qquad    X\equiv   1 + B^\dag B \ , \quad
Y\equiv\mathbf{1}_{N-1}  + B B^\dag \ ,  \label{X&Y}
\ee  
(known as the reducing matrix)  depends on the  $(N-1)$-component complex vector $B$,  the inhomogeneous coordinates of  ${\mathbbm {CP}}^{N-1}$.
The tension (\ref{tension}) obviously does not depend on the internal orientation  $\{B_i\}$.

That the vortex (internal)  moduli space is exactly a ${\mathbbm {CP}}^{N-1}$ in our model has been verified recently in \cite{Bolognesi:2014saa},\cite{Bolognesi:2015mpa} by studying these vortex solutions in the large-winding limit, where vortex configurations can be analytically determined, and accordingly all zero modes can be determined by following the analysis \`a la Nielsen-Olesen-Ambjorn \cite{Nielsen:1978rm}-\cite{Ambjorn:1989bd}.

In contrast with the standard nonAbelian vortices (which appear in theories with $g_r=0$ or $g_l=0$),  here 
the gauged vortex solutions with distinct moduli $\{B_i\}$ are related by global part of the $SU(N)$ gauge transformations.  
%In that sense all solutions related by  $U(B)$ are {\it  gauge equivalent}.  
The moduli can be promoted to collective coordinates if they are allowed to depend on $z$ and $t$.  The orthogonal parts of the collective coordinates can combine with each other, or with the $(z,t)$-independent parts of the fields, in gauge-invariant combinations.  As a result the corresponding oscillations do correspond to physical degrees of freedom, which are eaten in a kind of mini-Higgs mechanism.

%Nonetheless, as in any system in a broken local gauge symmetry, one must work in a fixed gauge. Thus if in a given gauge
%a vortex has a  particular orientation, e.g., (\ref{vortex}), its excitation along the orthogonal directions  (\ref{zeromodes}) represent physical modes  with nonvanishing energy.  {\bf{(Jarah: One does not really always need to work in a fixed gauge.  I would say that the point here is that the static solution and the excitation can be combined to create some gauge-invariant combinations, which is the reason that the excitations are physical.  The gauge-invariant combinations only depend on the orthogonal part of the fluctuation.)}}

Even in the case of vortices with constant $\{B_i\}$,  the {\it{relative}} orientations of multiple vortex systems do have observable effects and hence are physical \cite{Lo:1993hp},\cite{Bucher:1993jj},\cite{Bolognesi:2014saa}, \cite{Bolognesi:2015mpa}.   Consider two or more parallel
vortices with different orientations, $\{ B_i\}$.  Generalizing Eqs.~(\ref{loop0}-\ref{loopr}), particles carrying (for instance) fundamental charges with respect to $G=U_0(1)\times SU_L(N) \times SU_R(N)$, will experience various nonAbelian Aharonov-Bohm (AB) effects (``gauge transformations")  when encircling one of them,  $\psi \to \Gamma \, \psi$, 
\beq
\label{rotb}
\Gamma(B)&=&\left( e^{\frac{2 \pi i }{N}}, U(B)\left(\begin{array}{cc} e^{ 2 \pi i\frac{ g_{l}^2}{g^{\prime \, 2 }}\frac{N-1}{N}} &  \\ &  e^{- 2 \pi i\frac{ g_{l}^2  }{g^{\prime \, 2 }}\frac{1}{N}} {\mathbbm  1}_{N-1}  \end{array}\right) U(B)^{\dagger}, \right. \nonumber \\
&& \qquad  \left. U(B) \left(\begin{array}{cc} e^{ -2 \pi i\frac{g_{r}^2  }{g^{\prime \, 2 }}\frac{N-1}{N}} &  \\ &  e^{ 2 \pi i\frac{g_{r}^2  }{g^{\prime \, 2 }}\frac{1}{N}}    {\mathbbm 1}_{N-1}   \end{array}\right     ) U(B)^{\dagger}  \right)\ . 
\eeq
If a particle is in a generic representation of $G$,  it will experience an AB effect similar to the above  
but with appropriate charge and generators. 

If there are multiple vortices with orientations $\{B_{1}, B_{2}, \ldots \}$,  various closed paths encircling these vortices give rise to nonAbelian AB effects:   the gauge transformations experienced by a particle depend  on the order in which various vortices are circumnavigated. 
These and many other beautiful  features related to  such systems have been discussed in \cite{Lo:1993hp,Bucher:1993jj,Bolognesi:2014saa,Bolognesi:2015mpa}.

\section{Topological and geometric obstructions\label{sec:topological}}

In the vacuum our scalar condensate breaks the gauge symmetry $G$ down to a smaller gauge group $H$.  In the presence of a vortex the symmetry is broken yet further.  However, far from the vortex, the symmetry group $H$ is restored.  There are natural gauges, such as the regular gauge used above, in which the scalar condensate is position-dependent and therefore so is the embedding of $H$ in $G$.  The AB effect potentially makes the embedding of $H$ yet more complicated, as elements of $H$ do not generally commute with the AB phase.  

\subsection{Topological obstructions in general}

In various contexts it is useful to define a global symmetry corresponding to the gauge symmetry $H$.  The fact that the embedding of $H$ in $G$ is, in many gauges, position dependent means that such a global, continuous definition of the generators of $H$ may not exist for all values of the azimuthal angle $\theta$.    

Indeed there are many well-known cases where such an obstruction is known to exist.  For example, consider nonAbelian monopoles.  These are `t Hooft-Polyakov monopoles which preserve a nonAbelian symmetry group $H$ which is a subgroup of the ultraviolet gauge group $G$.  Consider a sphere $S^2$ which links a monopole.   It is known \cite{monopolonogo} that it is not possible to continuously define a set of generators of $H$ on this $S^2$.  The obstruction arises as follows.  The connection defines a trivial $G$ gauge bundle on $S^2$, as the gauge field is continuous and globally defined inside of the sphere.    The gauge symmetry $G$ is broken to $H$, and so one obtains an $H$ subbundle of the $G$ bundle.  However the $H$ bundle is nontrivial.  Indeed, it is the nontriviality of the $H$ bundle that gives `t Hooft-Polyakov monopoles their topological charge.  

This nontrivial bundle poses no problem for the existence of the monopole.  $H$ can be defined on northern and southern hemispheres of the $S^2$ and these hemispheres are related by a gauge transformation which corresponds to the generator of $\pi_1(H)$.  However when $H$ is nonAbelian, this gauge transformation acts nontrivially on any set of generators of $H$ and so implies that no set of generators of $H$ can be extended over the $S^2$.  This means that no global $H$ symmetry exists.  Colored dyons would be charged under such a global $H$ symmetry, and so the topological obstruction to a global definition of $H$ has the very physical consequence that no colored dyons exist.

A similar obstruction can occur in the case of vortices, as was discovered in Ref.~\cite{schwarz} which introduced the Alice string.  Again a symmetry group $G$ is broken to $H$.  In this case $H=U(1)$.  The vortex is linked by a circle $S^1$ and so the high energy theory is described by a trivial $G$ bundle on $S^1$ and the low energy theory by an $H$ bundle on $S^1$.  Again the $H$ bundle is nontrivial.  As the base space of the bundle is just a circle, the bundle can be trivialized on $\theta\in (0,2\pi)$ and so it is entirely characterized by the transition function when passing $\theta=2\pi$.

The Alice string is particularly exotic because a particle encircling the Alice string negates its electric charge.  Whenever particles encircling a vortex change the representation of $H$ under which they transform, the $H$ bundle is not principle.  Nonprinciple $U(1)$ bundles have transition functions in $Iso(U(1))$, the group of isometries of $U(1)$.  So each $U(1)$ bundle corresponds to an element of $Iso(U(1))$.  This is essentially an infinite dihedral group, it consists of multiplication by elements in $U(1)$ and the inversion of an element in $U(1)$.  Group multiplication by a fixed element of $U(1)$ yields the AB effect.  The total space of the $U(1)$ bundle is a torus and the multiplication by a fixed element simply means that the modulus of this torus is not purely imaginary, nonetheless the bundle is trivial.  On the other hand, the Alice string corresponds to the bundle in which the $U(1)$ fiber is inverted, corresponding to the negation of electric charge, when circumnavigating the vortex.  The total space of the bundle is the Klein bottle, it is not homeomorphic to the torus and the bundle is not trivial.  In particular, $H$ cannot be globally defined, since the generator of the Lie algebra of $H$ changes sign when circumnavigating the circle.

How can we classify such topological obstructions to the global definition of $H$?  The obstruction was a consequence of the nontrivial topology of the $H$ bundle over $S^1$ and indeed an obstruction implies a nontrivial $H$ bundle although the converse is not necessarily true.  These bundles are classified geometrically by an element of $Iso(H)$, the choice of monodromy when winding around the vortex.  In the case of gauged nonAbelian vortices, no particles change representations when encircling the vortex so the only elements of $Iso(H)$ which appear as monodromies are multiplication by elements of $H$.  Therefore our bundles over $S^1$ are in one to one correspondence with elements of $H$.  

Now for a {\it{topological}} obstruction we only are interested in an element of $H$ up to a continuous deformation.  Any two elements of $H$ in the same connected component of $H$ are related by a continuous deformation and so topologically $H$ bundles over $S^1$ are classified by $\pi_0(H)$, the set of connected components of $H$.  In particular, if the AB phase represents the trivial class in $\pi_0(H)$ then there is no topological obstruction to a global construction of $H$.

Similarly in the case of monopoles, nontrivial $H$ bundles over $S^2$ are classified by $\pi_1(H)$, although, as the original `t Hooft-Polyakov monopole case illustrates, a nontrivial bundle is not sufficient for a topological obstruction to the existence of a dyon, it is also necessary that the transition functions act nontrivially on the generators of the Lie algebra of $H$.

\subsection{Topological obstruction in our case?}

To determine whether or not there is a topological obstruction in our case, we will first need a more careful global definition of $G$ and $H$, paying particular attention to torsion elements, which are often responsible for topological obstructions.  There are three kinds of gauge fields, the photon carrying $U(1)_0$ and the gluons carrying the left and right $SU(N)$.  Thus the gauge group could in principle be as large as $U(1)\times SU(N)_l\times SU(N)_r$.  However the gauge symmetries which leave all of the fields invariant have no physical effect and so, to avoid confusion, should be quotiented out of the total gauge group.  There are four kinds of fields.  The three gauge fields and the scalar field $Q$.  All three gauge fields leave $U(1)_0$ invariant.  The left gauge fields $SU(N)_l$ are invariant under $U(1)_0\times SU(N)_r$ and also under their own center $\mathbb{Z}^l_N\subset SU(N)_l$.  Similarly the right gauge fields are invariant under $U(1)_0\times SU(N)_l\times \mathbb{Z}_N^r$.  

If these were the only fields, the total gauge symmetry would be $SU(N)_r/\mathbb{Z}_N^r\times SU(N)_l/\mathbb{Z}_N^l$.  However the scalar field $Q$ is charge one under $U(1)_0$, transforms in the fundamental of $SU(N)_l$ and the antifundamental of $SU(N)_r$.  This means that $SU(N)_l\times SU(N)_r$ acts freely on $Q$ except for the central $\mathbb{Z}_N$'s.  If one acts on $Q$ with $e^{ij/N}\in SU(N)_l$ and $e^{ik/N}\in SU(N)_r$ then the total effect will be to multiply $Q$ by $e^{i(j-k)/N}$.  Therefore one must quotient the gauge symmetry by the $\mathbb{Z}_N$ diagonal symmetry of $\mathbb{Z}_N^l\times\mathbb{Z}_N^r$, in other words by the elements for which $j=k$.  Including $U(1)_0$ makes the story slightly more complicated. Again $U(1)_0$ generally acts freely on $Q$ except for the $N$th roots of unity.  If one acts on $Q$ by $e^{im/N}\in U(1)_0$ as well as the central elements of $SU(N)_l\times SU(N)_r$ as above, then
\be
Q\longrightarrow e^{i(m+j-k)/N} Q
\ee
and so $Q$ is only invariant if
\be
m+j=k\ {\rm{mod}}\ N\ . \label{invar}
\ee
Therefore for any pair $(j,m)\in\mathbb{Z}_N^2$ there exists a $k$, given by (\ref{invar}), such that the corresponding gauge action leaves $Q$ invariant.  In conclusion, the total gauge group $G$ needs to be quotiented by this unphysical $\mathbb{Z}_N^2$, which leaves all of the fields in the theory invariant.  We then conclude
\be
G=\frac{U(1)_0\times SU(N)_l\times SU(N)_r}{\mathbb{Z}_N^2}\ .
\ee
If in addition one includes matter transforming in the fundamental of $SU(N)_l$ or $SU(N)_r$ then only a single $\mathbb{Z}_N$ should be quotiented, while with the inclusion of both there would be no quotient at all.

What about $H$?  This is the symmetry group left invariant by a vacuum value of $Q$.  This is not invariant under any rotation, so the $U(1)_0$ is eliminated.  Furthermore, as $Q$ is proportional to the identity, the $SU(N)_l$ and $SU(N)_r$ elements must be equal, leaving a single $SU(N)_{l+r}$.  The central $\mathbb{Z}_N\subset SU(N)_{l+r}$ consists of transformations such that $j=k$ and $m=0$ so they lie in the $\mathbb{Z}_N^2$ denominator in $G$ and so need to be quotiented out, leaving the projective unitary group
\be
H=\frac{SU(N)_{l+r}}{\mathbb{Z}_N}\ .
\ee  
Note that if matter transforming under only $SU(N)_l$ and/or $SU(N)_r$ is included then this $\mathbb{Z}_N$ is no longer quotiented out of $G$ and so $H=SU(N)_{l+r}$.  In either case, so long as all fields have integral $U(1)_0$ charge, $H$ is path connected and so $\pi_0(H)=0$.  Therefore no $H$ bundle over the circle can be nontrivial, so there is no topological obstruction to globally defining $H$. 

In fact, it is not difficult to explicitly construct elements of $H$ at arbitrary $\theta$.  $H$ is the symmetry group at large $r$, where Eq.~(\ref{vortex}) reduces to

\beq
Q=   \sqrt{\xi} \left(\begin{array}{cc}e^{i \theta} & 0 \\0 & {\bf 1}_{N-1}\end{array}\right)\ ,   
\eeq
corresponding to the gauge in which the modulus $B$ vanishes.  Elements of $H$ must preserve this form of $Q$.  Given an element $M$ of $SU(N)$, one can construct an element of $H$ as follows.  Let the $U(1)_0$ transformation be trivial.  Let the $SU(N)_l$ transformation be $M$ and let the $SU(N)_r$ transformation be
\beq
u_r={\rm{exp}}\left(-\frac{i\theta T_{N^2-1}}{NC_N}\right) M^\dagger {\rm{exp}}\left(\frac{i\theta T_{N^2-1}}{NC_N}\right).
\label{destra}
\eeq
This transformation preserves $Q$ because $Q$ times the first factor in the right hand side of (\ref{destra}) is proportional to the identity.

Summarizing, we have observed that $\pi_0(H)=0$ implies that all $H$ gauge bundles over a circle are trivial.  Therefore ours must be trivial and so no topological obstruction can exist to a global definition of $H$.  However if we allow matter with nonintegral $U(1)_0$ charges then $G$ and $H$ will change, becoming covers of the definitions above.  In such a case $\pi_0(H)$ is no longer necessarily trivial.  Nonetheless as $U(1)_0$ is abelian we do not expect this to lead to a topological obstruction in the definition of $H$.

What does this all mean physically?  In the monopole case, the fact that the `t Hooft-Polyakov monopole has no obstruction to a global definition of $H$ but its nonAbelian generalization does implies that the former can be modified to carry the electric charge but the latter cannot.  In the present case, the lack of a topological obstruction to the definition of $H$ means that the color electric charge of a perturbed vortex solution is well defined.  This color electric charge can be calculated by simply integrating the current multiplied by the $H$ generators described in (\ref{destra}) and traced.

\subsection{Geometric obstruction}

We have seen that there is no topological obstruction to the global definition of $H$ at all azimuthal angles $\theta$.  However the nonAbelian Aharonov-Bohm effect yields a closely related phenomenon: a geometric obstruction.  It is not possible to globally construct generators of $H$ which are covariantly constant with respect to the gauge field.

Recall that the scalar fields in the regular gauge have a nontrivial winding at spatial infinity 
\be
\label{wind}
\langle Q \rangle (\theta)=\sqrt{\xi}\,\,U\left(\begin{array}{cc}
e^{i\theta}&\\
&\mathbbm{1}\\
\end{array}\right) U^{-1}\, = \sqrt{\xi}\,\,e^{i\frac{\theta}{N}}\,\, e^{i\zeta_l\frac{\theta}{NC_N}T_{N^2-1}^{(U)}}\cdot \mathbbm{1}\cdot e^{i\zeta_r\frac{\theta}{NC_N}T_{N^2-1}^{(U)}}\ ,
\ee
 where the rotated $SU(N)$  generator is  given by $T_{N^2-1}^{(U)}=U\, T_{N^2-1}\,U^{\dagger}$.\footnote{We also recall that $\zeta_l \equiv  \tfrac{g_l^{2}}{g_l^2 + g_r^2}\ ,  \,\,   \zeta_r \equiv  \tfrac{g_r^{2}}{g_l^2 + g_r^2}\ . \label{zetalr}$}

A covariantly constant embedding of the unbroken symmetry group $H$ - the little group of   $\langle Q \rangle (\theta)$ -  inside the original symmetry group $G$ becomes $\theta$ dependent, and as a result some of the generators are not globally defined.

Let us rewrite (\ref{wind}) as  
\be\label{tw}
\langle Q \rangle (\theta)=\mathpzc{u}(\theta) \langle Q \rangle (0)\ ,
\ee
where $\mathpzc{u}(\theta)$ can be read off  from \eqref{wind} for the various simple factors in $G$:
\be
\label{g1}
\mathpzc{u}_1(\theta)= e^{i\frac{\theta}{N}}\ ,
\qquad 
\mathpzc{u}_l(\theta)=e^{i\zeta_l\frac{\theta}{NC_N}T^{(U)}_{N^2-1}}\ ,
\qquad 
\mathpzc{u}_r(\theta)=e^{-i\zeta_r\frac{\theta}{NC_N}T^{(U)}_{N^2-1}}\ .
\ee
This rewriting, equation $\eqref{tw}$, is  more adequate for a discussion valid in general gauge symmetry breaking systems, $G \to H$, with $\mathpzc{u}(\theta)\in G$, $\mathpzc{u}(0)=\mathbbm{1}$. 

In such systems, in order for the energy $\int  |D_iQ|^2$ to be finite,   the gauge field must approach   $A_i=-i\p_i \mathpzc{u}(\theta)\mathpzc{u}(\theta)^{-1}$ asymptotically. That is, $\mathpzc{u}(\theta)$ is determined by integrating the gauge field
\be
\mathpzc{u}(\theta)= \mathcal{P}\,e^{i\int_0^\theta A\cdot dl}\ ,
\ee
where the integral is computed along a circle at radial infinity. 

Let $T^a$ be a basis of generators for Lie($G$). Define the Lie algebra automorphism
\be
\label{aut}
T^a\to T^a(\theta)=\mathpzc{u}(\theta)T^a \mathpzc{u}(\theta)^{-1}\ .
\ee    
A covariantly constant embedding of the unbroken symmetry group $H$ inside $G$ varies with $\theta$:
\be  T^a(\theta)  \langle Q \rangle (\theta) =0\  \qquad  {\rm  if}  \qquad   T_a \,  \langle Q \rangle (0)=0\ . \ee 
  In general,  after a full circuit
\be
T^a(2\pi)=\mathpzc{u}(2\pi)T^a\mathpzc{u}(2\pi)^{-1}=O^{ab}T^b\ ,
\ee
where $O$ must be a real orthogonal matrix to preserve Hermiticity and normalization of the generators with respect to the invariant inner product $(T^a,T^b):=\Tr(T^a T^b)$. By a basis change in Lie($H$) $O$ can always be diagonalized (every orthogonal matrix is unitarily diagonalizable over $\mathbb C$). The diagonal basis will involve in general complex linear combinations of the original generators. However, since $O$ is real its eigenvalues come in conjugate pairs $\lambda$ and $\lambda^*$ and corresponding eigenvectors $v$ and $v^\dagger$,      i.e.,    $v+v^\dagger$ and $i(v-v^\dagger)$. In such a basis
\be
\label{to}
T^a(2\pi)=e^{2\pi i\xi^a}T^a\ .
\ee 
The covariantly constant generators for which $\xi^a=0$ are globally well-defined and generate the unbroken symmetry group $\widetilde H \subset H$. 
% In other words, $\widetilde H$ coincides with the centralizer of $\mathpzc{u}(2\pi)$. 
 The generators for which $\xi^a\neq 0$ are not. 

\section{Zeromode excitations \label{sec:excitation}}

Let us now allow the vortex orientation zeromodes   $\{B_i\}$ to fluctuate.    If  $\{B_i\}$ are allowed to depend only on $\{z, t\}$ 
 %that the whole vortex solution fluctuates rigidly,  
  the field equations  (\ref{BPSequations2})-(\ref{BPSequations4})  containing the $x$ and $y$ derivatives and the corresponding gauge fields components are unmodified.  The other field equations however are modified, and in order to describe the zero mode excitations one must take into account the correct  response of the gauge fields to the $\{z, t\}$ modulation of the vortex orientation. 
The new gauge components,  induced by the %charge and currents along the vortex direction,  are just 
the nonAbelian Gauss and Biot-Savart effects,  satisfy the following equations of motion:
   \be
\D_i F_{i\alpha}^{(l)}=ig_l T^a\Tr \,   [\, Q^\dagger T^a\D_\alpha Q-  (\D_\alpha Q)^{\dagger}  T^a  Q\,] \,  \label{gauss1}
\ee
and
\be
\D_i F_{i\alpha}^{(r)}=-ig_r T^a\Tr  \, [\,T^aQ^\dagger\D_\alpha Q-   (\D_\alpha Q)^{\dagger} Q  T^a \, ]\,\label{gauss2}
\ee
($\alpha=z,t$ ;   $i=x,y$).

It turns out that the form of the new gauge field components consistent with these equations is given by the following Ansatz
\be
a_\alpha=0\ ,
\ee
\be 
A_\alpha^{(l)}=\rho_l W_{\alpha}+\eta_l V_{\alpha}\ ,  \label{color1}
\qquad  
A_\alpha^{(r)}=\rho_r   W_{\alpha}+\eta_r V_\alpha\ , 
\ee
where 
\be\begin{split}
&W_\alpha \equiv   i \, \p_\alpha T^{(U)} T^{(U)} \ , \qquad  V_{\alpha}\equiv \p_\alpha T^{(U)} \ ,  \\ 
&  T^{(U)}  =  U T U^{\dagger}\ ,  \qquad\qquad   T=  \left(\begin{array}{cc}1 &  \\ & -{\bf 1}_{N-1}\end{array}\right)\ .     \label{color3}
\end{split}\ee 
%$U=U(B(x_{\alpha}))$ is slowly varying with $t,z$. 
 The profile functions $\rho(r,\theta)$ and $\eta(r,\theta)$, as we are going to see next,  satisfy the equations that follow from the insertion of the Ansatz into equations (\ref{gauss1}) and (\ref{gauss2}).

 For the  standard, ungauged, nonAbelian vortices (with $g_{r}=0$ or $g_{l}=0$),  it was convenient to develop the effective action using the vortex solution in  the singular gauge, the one  in which the scalar fields do not wind \cite{Auzzi:2003fs}.   
For the gauge nonAbelian vortex instead the use of the singular gauge is somewhat subtle, as the gauge fields in such a gauge develops Dirac sheet singularities \cite{Bolognesi:2014saa}.    
Below we shall work in the regular gauge, Eq.~(\ref{vortex}) - Eq.~(\ref{Ar}). We shall also see that  result obtained for the ungauged vortex in the singular gauge is correctly reproduced.

The price that we pay for using the regular gauge is that  the equations of motion take a more complicated form as the background and quantum fields depend upon the azimuthal angle.   The  color structure of the gauge fields   (\ref{color1})  is also richer than in the case of the 
standard nonAbelian vortices, where the only color structure (in the singular gauge) was 
 \beq   W_{\alpha} = i \, \p_\alpha T^{(U)} T^{(U)}=
%   i\, U \, \{  U^{\dagger} \de_{\alpha} U  -  T\,  U^{\dagger} \de_{\alpha}  U \, T  \}   U^{\dagger} = 
2 i\,   U \, (U^{\dagger} \de_{\alpha} U)_{\perp} U^{\dagger}\;, \label{form}
\eeq
where 
\beq   (U^{\dagger} \de_{\alpha} U)_{\perp} \equiv   \frac{1}{2} \{  U^{\dagger} \de_{\alpha} U  -  T\,  U^{\dagger} \de_{\alpha}  U \, T  \} \ 
\eeq
is the Delduc-Valent projection \cite{Delduc} on the Nambu-Goldstone direction in the tangent space. The form (\ref{form}) means that the massless excitations correspond to the Nambu-Goldstone modes, whose direction must be kept orthogonal to the  "rotating  background'',  $U Q U^{\dagger}$ \cite{Gudnason:2010rm}.
The more general form of the gauge field Ansatz (\ref{color1}) contains an additional term that  arises from the gauge transformation to the regular gauge  (see Eq.~(\ref{seebelow}) below).  
The strongest justification for the Ansatz, however, comes from the fact it allows one to resolve the color structure of the equations in a closed form (\ref{gauss1}), (\ref{gauss2}).

To lowest order the excitations above the static, constant tension vortex are described by the effective action 
\be
S_{eff}=\int \,dt  d^{3}x\,  [\Tr(F_{i\alpha}^{(l)}F^{(l)\,\,\alpha}_{\,\,\,\,\,\,i})+\Tr(F_{i\alpha}^{(r)}F^{(r)\,\,\alpha}_{\,\,\,\,\,\,i})+\Tr|D_\alpha Q|^2]\ .
\label{Action}\ee
We neglect here the terms coming from $F_{\alpha\beta}^{(l/r) 2}$ which would generate higher derivative terms in the effective action; we shall come back later to discuss the validity of this approximation. 
By inserting the Ansatz, a straightforward calculation \cite{L.S.}  yields 
\be  \mathcal{S}_{eff} =   {\cal I} \,  \int \, dz \,dt  \,  \Tr(\p_\alpha T^{(U)})^2\,,     \label{effectiveAc} \ee
 where    ($\sigma_l \equiv 1+2g_l\rho_l\,
$,    $\sigma_r \equiv 1+2g_r \rho_r \,
$)  
 \be \begin{split}
\label{eff}
 {\cal I}   &=\int \,d\theta\,dr\, r \bigg[(\p_r\rho_l)^2+\frac{1}{r^2}(\p_\theta \rho_l)^2+(\p_r\eta_l)^2+\frac{1}{r^2}(\p_\theta \eta_l)^2+\\
&\ \ \ +\left(\frac{g_l}{2g'^2}\right)^2\left(\frac{1-f_{NA}}{r}\right)^2(\sigma^2_l+4g^2_l\eta^2_l)-\left(\frac{g_l}{g'^2}\right)\left(\frac{1-f_{NA}}{r^2}\right)(\sigma_l\p_\theta\eta_l-2g_l\eta_l\p_\theta\rho_l)+\\
&\ \ \ +(\p_r\rho_r)^2+\frac{1}{r^2}(\p_\theta \rho_r)^2+(\p_r\eta_r)^2+\frac{1}{r^2}(\p_\theta \eta_r)^2+\\
&\ \ \ +\left(\frac{g_r}{2g'^2}\right)^2\left(\frac{1-f_{NA}}{r}\right)^2(\sigma^2_r+4g^2_r\eta^2_r)+\left(\frac{g_r}{g'^2}\right)\left(\frac{1-f_{NA}}{r^2}\right)(\sigma_r\p_\theta\eta_r-2g_r\eta_r\p_\theta\rho_r)+\\
&\ \ \ +\frac{Q_1^2+Q_2^2-2Q_1Q_2\cos\theta}{4}[(1+g_l\rho_l+g_r\rho_r)^2+(g_l\eta_l+g_r\eta_r)^2]+\\
&\ \ \ +\frac{Q_1^2+Q_2^2+2Q_1Q_2\cos\theta}{4}[(g_l\eta_l-g_r\eta_r)^2+(g_l\rho_l-g_r\rho_r)^2]+\\
&\ \ \ + Q_1Q_2\,\sin\theta(g_r\eta_r\sigma_l-g_l\eta_l\sigma_r)\bigg]\ , 
\end{split}\ee
and 
\be  \int dz dt  \,  \Tr(\p_\alpha T^{(U)})^2 =   8\,  \int dz dt  \,   X^{-1} \partial_{\alpha}B^{\dagger}  Y^{-1}  \partial_{\alpha}  B\ 
\ee
is the standard ${\mathbbm {CP}}^{N-1}$ sigma model action. 
The equations for the  profile functions $\rho$ and $\eta$ can be determined by minimizing ${\cal I}$. Alternatively they can be derived directly from the equations of motion (\ref{gauss1}) and (\ref{gauss2}).

Note that,  in spite of the fact that terms with two different color structures $W_\alpha=i \p_\alpha T^{(U)} T^{(U)} $ and $V_{\alpha}=\p_\alpha T^{(U)} $ appear in various contributions,  the total action is simply proportional  to $\Tr(\p_\alpha T^{(U)})^2 $,  due to the following identities: 
\be   \Tr  \, W_{\alpha}^{2}=   \Tr  \, V_{\alpha}^{2}=4N\,    X^{-1} \partial_{\alpha}B^{\dagger}  Y^{-1}  \partial_{\alpha}  B\ ,
\qquad \Tr  \, W_{\alpha}V_{\alpha}=0\ . 
\ee

\subsection{Equations for $\rho$ and $\eta$}

By minimizing ${\cal I}$ with respect to  $\rho_r$,  $\eta_r$, $\rho_l$ and  $\eta_l$,  {\it given} the other functions $Q_{1,2}$, $f_0$, $f_{NA}$ fixed, one finds the following four coupled equations 
\be \label{p1}  \begin{cases}
\begin{split}
&\frac{1}{g_l^2}\Delta \eta_l-\frac{2}{g'^2}\frac{1-f_{NA}}{r^2}\p_\theta \rho_l-\left(\frac{g_l}{g'^2}\right)^2\left(\frac{1-f_{NA}}{r}\right)^2\eta_l-\frac{Q_1^2+Q_2^2}{2}\eta_l+\frac{Q_1Q_2\,\sin\theta}{2g_l}\sigma_r+\\&+\frac{Q_1Q_2\,\cos\theta}{g_l}g_r\eta_r=0\ ,
\end{split}    
\\
\begin{split}
&\frac{1}{g_l^2}\Delta \rho_l+\frac{2}{g'^2}\frac{1-f_{NA}}{r^2}\p_\theta\eta_l-\frac{g_l}{2g'^4}\left(\frac{1-f_{NA}}{r}\right)^2-\left(\frac{g_l}{g'^2}\right)^2\left(\frac{1-f_{NA}}{r}\right)^2\rho_l+\\&-\frac{Q_1^2+Q_2^2-2Q_1Q_2\cos\theta}{4g_l}-\frac{Q_1^2+Q_2^2}{2}\rho_l+\frac{Q_1Q_2\,\cos\theta}{g_l}g_r\rho_r-\frac{Q_1Q_2\,\sin\theta}{g_l}g_r\eta_r=0\ .
\end{split}
\end{cases}   \ee

\be\label{p2}\begin{cases}
\begin{split}
&\frac{1}{g_r^2}\Delta \eta_r+\frac{2}{g'^2}\frac{1-f_{NA}}{r^2}\p_\theta \rho_r-\left(\frac{g_r}{g'^2}\right)^2\left(\frac{1-f_{NA}}{r}\right)^2\eta_r-\frac{Q_1^2+Q_2^2}{2}\eta_r-\frac{Q_1Q_2\,\sin\theta}{2g_r}\sigma_l+\\&+\frac{Q_1Q_2\,\cos\theta}{g_r}g_l\eta_l=0\ ,
\end{split}
\\
\begin{split}
&\frac{1}{g_r^2}\Delta \rho_r-\frac{2}{g'^2}\frac{1-f_{NA}}{r^2}\p_\theta\eta_r-\frac{g_r}{2g'^4}\left(\frac{1-f_{NA}}{r}\right)^2-\left(\frac{g_r}{g'^2}\right)^2\left(\frac{1-f_{NA}}{r}\right)^2\rho_r+\\&-\frac{Q_1^2+Q_2^2-2Q_1Q_2\cos\theta}{4g_r}-\frac{Q_1^2+Q_2^2}{2}\rho_r+\frac{Q_1Q_2\,\cos\theta}{g_r}g_l\rho_l+\frac{Q_1Q_2\,\sin\theta}{g_r}g_l\eta_l=0\ .
\end{split}
\end{cases}   \ee

It can be verified that the same equations follow directly from the Gauss equations, after factoring out the common color structures  $W_{\alpha}$ and $V_{\alpha}$.    
   
%%%%%%%%%%%%%%%%%%%%%%%%%%%%%%%%%%%%%%%%%%%%%%%%%%%%%%%%%%%%%%%%%%%%%%%%%%%%%%%%%%%%%%%%%%%%%%%%%%%%%%%%%%%%%%%%%%%%%%%%

\subsection{Global  ($g_r=0$) nonAbelian vortex \label{sec:check}}

A nontrivial check of the equations found above is provided by the  consideration of the ungauged vortex case. 
After setting the right gauge coupling to zero $g_r=0$ and $\eta_r=\rho_r=0$, equations (\ref{p1}) reduce to   ($g_l \equiv g$, $\eta_l \equiv \eta$ and $\rho_l \equiv \rho$)
\be\begin{cases}
\begin{split}
&\Delta\eta-2\,\frac{1-f_{NA}}{r^2}\p_{\theta}\rho-\left(\frac{1-f_{NA}}{r}\right)^2\eta+\frac{Q_1Q_2\,\sin\theta}{2}g-\frac{Q_1^2+Q_2^2}{2}g^2\eta=0\ ,
\end{split}
\\
\begin{split}
&\Delta\rho+2\,\frac{1-f_{NA}}{r^2}\p_{\theta}\eta-\frac{1}{2g}\left(\frac{1-f_{NA}}{r}\right)^2-\left(\frac{1-f_{NA}}{r}\right)^2\rho-\frac{Q_1^2+Q_2^2-2Q_1Q_2\cos\theta}{4}g   \\&-\frac{Q_1^2+Q_2^2}{2}g^2\rho=0\ .
\end{split}
\end{cases}    \label{global}\ee
In order to compare them to the equations studied earlier \cite{Auzzi:2003fs}-\cite{Gudnason:2010rm},   it is necessary to gauge transform form the singular to the regular gauge.  
In the singular gauge the gauge field $A_{\alpha}^{(s)}$  is given by 
\be
A^{(s)}_\alpha=  \rho^{(s)}  \, W_\alpha\ . \ee
The gauge transformation from the singular to regular gauge is achieved by  
\be
u =e^{i\theta T^{(U)}_{N^2-1}/NC_N}\ .  \label{just}
\ee
That is: 
\be
A_\alpha= u A^{(s)}_\alpha u^{-1}-\frac{i}{g}\p_{\alpha} u  u^{-1}=\frac{1}{2g}(\cos\theta\,\sigma^{(s)} -1)\,W_\alpha+\frac{1}{2g}(\sin\theta\,\sigma^{(s)})\,V_\alpha\ ,    \label{seebelow}
\ee
where  $\sigma^{(s)} \equiv  1 + 2 g \rho^{(s)}.$
Apart from  an irrelevant  $U_0(1)$ transformation $ e^{i \theta /N}$,    (\ref{just})   is just a gauge transformation
\be    u=    U\,  \left(\begin{array}{cc}e^{ i \theta} & 0 \\0 & {\bf 1}_{N-1}\end{array}\right)  \, U^{\dagger} \ ,  \label{justagauge}
\ee
which winds the scalar fields once.
A useful relation is 
\be
\frac{T_{N^2-1}}{NC_N}=\frac{N-2}{2N}\,\mathbbm{1}_N+\frac{T}{2}\ , \qquad  \frac{T^{(U)}_{N^2-1}}{NC_N}=\frac{N-2}{2N}\,\mathbbm{1}_N+\frac{T^{(U)}}{2}\ .
\ee
The profile functions are accordingly transformed as 
\be\begin{cases}
\begin{split}
\rho^{(s)} \to   \rho=\frac{1}{2g}(\cos\theta\,\sigma^{(s)}-1)\ ,
\end{split}
\\
\begin{split}
\eta^{(s)} (=0)  \to      \eta=\frac{1}{2g}(\sin\theta\,\sigma^{(s)})\ .
\end{split}
\end{cases}    \label{gaugetrprof} \ee
This gauge transformation   can be expressed in a more elegant form  by introducing the complex combination of the  profile functions
\be \psi  \equiv    \sigma +  2 i g  \eta\;,\qquad   \sigma =  1 + 2 g \rho 
\;; \ee
\be  \psi^{(s)} =   \sigma^{(s)}\;,  \qquad   \sigma^{(s)} =  1 + 2 g \rho^{(s)}\;.
\ee
Eq.~(\ref{gaugetrprof}) then becomes simply
\be   \psi^{(s)} \to     \psi = e^{i \theta}  \psi^{(s)}\;.        \label{gtpsi}
\ee

At this point it is a simple matter to verify  that  our equations in the regular gauge give the same profile function
(for the $g_r=0$ theory)  known from the earlier studies. 
We write the equations  (\ref{global})  in terms of $\psi$ as  a  single complex equation:
\be  \Delta \psi - 2 i  \frac{ 1- f_{NA}}{r^2}  \p_{\theta} \psi -   \left(\frac{1-f_{NA}}{r}\right)^2  \psi   + g^2\, Q_1 Q_2  e^{i \theta}  - \frac{g^2}{2}
(Q_1^2 + Q_2^2 ) \psi =0\ .
\ee
By substituting  (\ref{gtpsi})   into this equation 
%and recalling that in the singular gauge
%\[  \psi^{(s)}=  \sigma^{(s)}(r)=  1 + 2 g \rho^{(s)}(r)\ , 
%\]
one gets, after some simple algebra,
\be 
\frac{1}{r} \p_r \left(r \p_r \, \psi^{(s)}\right)  -     \left( \frac{f_{NA}}{r}\right)^2   \,\psi^{(s)}   -  g^2 \left[   \frac{Q_1^2 +  Q_2^2 }{2}    \,\psi^{(s)} -  Q_1 Q_2 \right]=0\;.
\ee
This is precisely the equation for  the profile function    $  \psi^{(s)}  =  \sigma^{(s)}= 1 + 2 g \rho^{(s)}$ found earlier
 \cite{Auzzi:2003fs}-\cite{Gudnason:2010rm} in the singular gauge,  whose solution is given by 
 \be    \psi^{(s)}  =   \sigma^{(s)}=  \frac{Q_1}{Q_2}\;,
 \ee
 as can be shown by using the BPS equations for  $Q_1, Q_2, f$ and $f_{NA}$.
The result for $\psi$    in the regular gauge is then
\be
\psi = e^{i \theta} \,    \frac{Q_1}{Q_2} \;.  \label{ungauged}
\ee

%%%%%%%%%%%%%%%%%%%%%%%%%%%%%%%%%%%%%%%%%%%%%%%%%%%%%%%%%%%%%%%%%%%%%%%%%%%%%%%%%%%%%%%%%%%%%%%%%%%%%%%%%%%%%%%%%%%%%%%%

\section{Solution of Gauss's Equations and Vortex Excitation Energy \label{sec:energy}}

To study the solutions of equations (\ref{p1}) and  (\ref{p2}) and to obtain the associated  excitation energy,  it is 
convenient to use the complex combination of the profile functions $(\rho, \eta)$ already introduced in Section~\ref{sec:check},  this time both for the left and right fields: 
\beq
&&\psi_l \equiv \sigma_l+2ig_l\eta_l\ , \qquad  \ \,  (\sigma_l= 1 + 2 g_l \rho_l)\ , \nonumber \\
&&\psi_r  \equiv  \sigma_r+2ig_r\eta_r\ ,  \qquad   (\sigma_r= 1 + 2 g_r \rho_r)\ . 
\eeq
Then equations $\eqref{p1}$ and $\eqref{p2}$ reduce to two complex equations
\be
\label{ee}
\Delta \psi_l-2\zeta_l\frac{1-f_{NA}}{r^2}\p_\theta(i\psi_l)-\left(\frac{\zeta_l(1-f_{NA})}{r}\right)^2\psi_l-g_l^2\left(\frac{Q_1^2+Q_2^2}{2}\psi_l-Q_1Q_2e^{i\theta}\psi_r\right)=0\ ,
\ee
\be
\label{ef}
\Delta \psi_r+2\zeta_r\frac{1-f_{NA}}{r^2}\p_\theta(i\psi_r)-\left(\frac{\zeta_r(1-f_{NA})}{r}\right)^2\psi_r-g_r^2\left(\frac{Q_1^2+Q_2^2}{2}\psi_r-Q_1Q_2e^{-i\theta}\psi_l\right)=0\ .
\ee
%and their complex conjugate. 

\subsection{Partial-waves and asymptote  \label{sec:asymptotes}}

These equations can be solved via the partial wave decomposition:
\be \label{Ansatz}
\psi_l=\sum_{m\in {\mathbbm Z}} \phi_l^m e^{im\theta}\ ,\qquad \ \ 
\psi_r=\sum_{m\in {\mathbbm Z}}  \phi_r^m e^{i(m-1)\theta}\ . 
\ee
Note the particular way that the left and right fields are correlated in angular momentum, reflecting the minimum winding of the vortex. 
The $\theta$ dependence in fact  drops out of  equations $\eqref{ee}$ and  $\eqref{ef}$  and one gets an infinite tower of 
  pairs of $\left\{\phi_l^m(r),  \phi_r^m(r)\right\}$ decoupled from each other and satisfying\footnote{$\Delta_r$ denotes the radial part of the two dimensional Laplacian:
$
\Delta_ra_\alpha=\tfrac{1}{r}\p_r(r\p_r a_\alpha)\ .
$} 
\be
\label{e1}
\Delta_r \phi_l^m-\left(\frac{m-\zeta_l A}{r}\right)^2\phi_l^m-g_l^2\left(\frac{Q_1^2+Q_2^2}{2}\phi_l^m-Q_1 Q_2\phi_r^m\right)=0\ , \ \ \ 
\ee
\be
\label{e2}
\ \ \ \ \ \ \ \, \Delta_r \phi_r^m-\left(\frac{m-1+\zeta_r A}{r}\right)^2\phi_r^m-g_r^2\left(\frac{Q_1^2+Q_2^2}{2}\phi_r^m-Q_1 Q_2\phi_l^m\right)=0\ ,
\ee
 with
 \be A(r)=1-f_{NA}(r)  \ , \qquad    A(0) =0\ , \quad  A(\infty)=1\ .  \ee
 Taking the difference of $\eqref{e1}$ and $\eqref{e2}$, one can prove that  $\phi_l^m\to\phi_r^m$ exponentially fast at spatial infinity. 
The asymptotic form of equation $\eqref{e1}$ for $r\to\infty$, where $Q_1,Q_2\to\sqrt{\xi}$, $A\to1$ and $\phi_l^m\to\phi_r^m$, is
\be
\label{e3}
\left(\p_r^2+\frac{1}{r}\p_r\right)\phi_l^m-\left(\frac{m-\zeta_l}{r}\right)^2\phi_l^m=0\ .
\ee
This equation  has two independent solutions:
\be  \phi_l^m\propto r^{\pm(m-\zeta_l)}\;,   \qquad \qquad {\rm if} \ \  m\neq\zeta_l \label{asympt} \ , \ee
or 
\be  \  \phi_l^m\propto\ln(r)\;, \quad  {\rm const.}\ \ \  \qquad {\rm if} \ \  m=\zeta_l  \ .
\ee
Similarly, the asymptotic form of equation $\eqref{e2}$ 
for  $\phi_r^m$  is
\be
\left(\p_r^2+\frac{1}{r}\p_r\right)\phi_r^m-\left(\frac{m-1+\zeta_r}{r}\right)^2\phi_r^m=0\ ,
\ee
which is the same as equation $\eqref{e3}$ since $\zeta_l+\zeta_r=1$.

Near  the vortex core ($r=0$),  where $Q_1\to0$ and $A\to 0$,   Eq.~\eqref{e1} behaves as
\be
\left(\p_r^2+\frac{1}{r}\p_r\right)\phi_l^m-\left(\frac{m}{r}\right)^2\phi_l^m=0\ .
\ee
This equation has  two independent solutions: 
\be 
\label{divphil}
\phi_l^m\propto r^{\pm m}\    \qquad \qquad \ \ {\rm if} \ \ m\neq 0  \ ,\ee
or 
\be  \   \  \phi_l^m\propto\log r,  \quad {\rm  const.}\ \  \qquad {\rm if} \ \  m=0  \  . \ee
 Similarly, from $\eqref{e2}$:
\be 
\label{divphir}
\phi_r^m\propto r^{\pm (m-1)}\    \qquad \qquad \ \ {\rm if} \ \ m\neq 0  \ ,\ee
or 
\be  \  \phi_r^m\propto\log r,  \quad{\rm  const.}\ \ \ \ \  \qquad {\rm if} \ \  m=0  \  . \ee
For the $m=0$ wave, the  solution  behaving as $\log  r$ at the origin  is excluded by the regularity requirement, and so are 
the negative-power solutions for $m \ne0$.

\subsection{Exact solution} 

Equations (\ref{ee}) and (\ref{ef}) can be solved exactly once the vortex profile functions are known. 
Define the function $\varphi=\varphi(r)$ as follows: 
\begin{eqnarray}
 \varphi\equiv -\log Q_1(r)+\log Q_2(r) +\log r\ ,\qquad
e^{-\varphi}=\frac{Q_1}{r Q_2}\ . 
\end{eqnarray}
 $\varphi$ is regular at $r=0$ and 
behaves as  $\varphi \simeq \log r$ at large $r$.  
With $f,f_{NA}, Q_1,Q_2$ defined in Eq.(\ref{bpsvortexuno})-(\ref{bpsvortextre}), $\varphi(r)$  satisfies the differential equations
\begin{eqnarray}
 \frac{1-f_{NA}}r= \varphi'\ ,\qquad 
\frac1r (r\varphi)'=-\frac{{g'}^2}2\left(Q_1^2-Q_2^2\right)\ . \label{thfirst}
\end{eqnarray}
The properties of the function $\varphi$  have been studied in detail in  \cite{Deeply}.

By using  the first equation of Eq.(\ref{thfirst}),
Eq.(\ref{ee}) and Eq.(\ref{ef}) can be rewritten as
\begin{eqnarray}\label{kore}
 \left\{\Delta -2\zeta_l \frac{\varphi'}r\, i\partial_\theta-(\zeta_l\varphi')^2 
-\frac{\zeta_l {g'}^2}2\left(Q_1^2+Q_2^2\right)\right\}\psi_l
+\zeta_l {g'}^2Q_1Q_2e^{i\theta}\, \psi_r&=&0\ ,\nn
 \left\{\Delta +2\zeta_r \frac{\varphi'}r\, i\partial_\theta-(\zeta_r\varphi')^2 
-\frac{\zeta_r {g'}^2}2\left(Q_1^2+Q_2^2\right)\right\}\psi_r
+\zeta_r {g'}^2 Q_1Q_2e^{-i\theta}\, \psi_l&=&0\ . \nn  
\end{eqnarray}
Note the following $\mathbb Z_2$ symmetry
\begin{eqnarray}
 (\psi_l, \zeta_l) \ \  \longleftrightarrow \ \  (\psi_r^*,\zeta_r)\ .
\end{eqnarray}
In order to simplify the equations we set
\be  \psi_l=e^{-\zeta_l \varphi}\widetilde \psi_l\ , \qquad
\psi_r=e^{\zeta_r \varphi}\widetilde \psi_r\ ,\ee
and we use a complex formulation. Eqs.(\ref{kore}) then become 
\begin{eqnarray}
   \left\{\Delta -4\zeta_l \varphi'\, \frac{\bar z}{r} \partial_{\bar z} 
-g^2_lQ_2^2 \right\}\widetilde \psi_l
+g^2_l Q_2^2 \,z\, \widetilde \psi_r&=&0\ ,\nn
 \left\{\Delta  +4\zeta_r \varphi'\, \frac{\bar z}{r} \partial_{\bar z} 
-g^2_r Q_1^2\right\}\widetilde \psi_r
+g_r^2 Q_1^2 z^{-1}\,\widetilde \psi_l&=&0\ .  \label{satisfy}
\end{eqnarray}
It can now be  easily  seen that the holomorphic functions  $\widetilde \psi_l, \widetilde \psi_r$ that satisfy
\begin{eqnarray}
\partial_{\bar z}\widetilde \psi_l=0\ , \qquad 
\partial_{\bar z}\widetilde \psi_r=0\ ,\qquad  {\rm and}\quad 
\widetilde \psi_l=z \, \widetilde \psi_r\ 
\end{eqnarray}  
solve (\ref{satisfy})   ($\Delta \propto \de_z \de_{\bar z}$). 
Using the $\mathbb Z_2$ symmetry, we find another set of  solutions.  
The general solution to Eq.(\ref{ee}) and Eq.(\ref{ef}) is therefore given by the linear combinations 
\begin{eqnarray}
 \psi_l&=&e^{-\zeta_l\varphi}\, z\, \chi_r(z)+e^{\zeta_l \varphi}\, \bar
  \chi_l(\bar z)\ ,\nn
\psi_r&=&e^{\zeta_r\varphi}\, \chi_r(z)+e^{-\zeta_r \varphi}\,
\bar z\,\bar
  \chi_l(\bar z)\ , \label{exactsolution}
\end{eqnarray}
in terms of two arbitrary holomorphic  functions $\chi_r(z)$ and $ \chi_l(z)$. At large $r$, they behave as
\begin{eqnarray}
  \psi_l&\simeq& e^{i\theta}\,r^{\zeta_r}\, \chi_r(z)
+r^{\zeta_l}\, \bar
  \chi_l(\bar z)\simeq e^{i\theta}\psi_r\ ,\nn
\psi_r&\simeq &r^{\zeta_r}\, \chi_r(z)+
e^{-i\theta}r^{\zeta_l} \,\bar
  \chi_l(\bar z)\ .
\end{eqnarray} 
  The solution of the minimum excitation energy ($m=1$ wave for $g_r<g_l$,  see the next subsection)  corresponds to the particular choice above,
\begin{eqnarray}
 \chi_r(z)={\rm const}.\;;   \qquad {\bar \chi}_l({\bar z})=0\;.
\end{eqnarray}
 The holomorphic and antiholomorphic terms correspond to the positive and negative angular momenta respectively in the partial wave decomposition, 
  (\ref{Ansatz}).

For the  special choice of the $U_0(1)$ coupling, 
\begin{eqnarray}
\label{simplifycouplings}
 2 N g_0^2={g'}^2   \equiv g_r^2+g_l^2\ ,
\end{eqnarray}
 Eqs (2.17)-(2.19)   
 become simply  
\begin{eqnarray}
\label{simplifyvortex}
f_{NA}=f=1-r \varphi'\,;\qquad  Q_1=\sqrt{\xi} \,r  e^{-\varphi}\,;\qquad Q_2= \sqrt{\xi} \ ,
\end{eqnarray}
and the function $\varphi=\varphi(r)$ in this case coincides with the solution of Taubes' equation
\begin{eqnarray}
 \frac1r(r \varphi)'=\frac{m_0^2}2(1-e^{-2\varphi}r^2)\ ,
\end{eqnarray}
with
\begin{eqnarray}
 \lim_{r\to 0} r\varphi' =0\ ,\qquad   \lim_{r\to \infty} (\varphi-\log r)=0\ ,
\end{eqnarray}
and with  $m_0^2\equiv g'^2 \xi=2 N g_0^2   \xi.$

\subsection{Divergences of the energy  \label{divergences}}

To study  the excitation  energy  we consider the integral   ${\cal I}$ in (\ref{eff})  which appears in front of the ${\mathbbm {CP}}^{N-1}$ action  in  $\eqref{effectiveAc}$.  In terms of $\psi_l$ and $\psi_r$ it simplifies considerably:
\be\label{ci}\begin{split}
\mathcal{I}=\int \,d\theta\,dr\, r &\bigg[\frac{1}{4g^2_l}\bigg(|\p_r\psi_l|^2+\frac{1}{r^2}|\p_\theta\psi_l|^2+\bigg(\frac{A\zeta_l}{r}\bigg)^2|\psi_l|^2-2\zeta_l\frac{A}{r^2}\Im(\psi^*_l\p_\theta\psi_l)\bigg)+\\
&+\frac{1}{4g^2_r}\bigg(|\p_r\psi_r|^2+\frac{1}{r^2}|\p_\theta\psi_r|^2+\bigg(\frac{A\zeta_r}{r}\bigg)^2|\psi_r|^2+2\zeta_r\frac{A}{r^2}\Im(\psi^*_r\p_\theta\psi_r)\bigg)+\\
&+\frac{Q_1^2+Q_2^2}{8}(|\psi_l|^2+|\psi_r|^2)-\frac{Q_1Q_2}{2}\Re(\psi_l\psi^*_re^{-i\theta})\bigg]\ .
\end{split}
\ee
First of all, we note that this expression is positive semidefinite:
\be
\label{brok}
\begin{split}
\mathcal{I}\geq \int \,d\theta\,dr\, r &\bigg[\frac{1}{4g^2_l}\bigg(|\p_r\psi_l|^2+\frac{1}{r^2}\left(|\p_\theta\psi_l|-A\zeta_l|\psi_l|\right)^2\bigg)+\\
&+\frac{1}{4g^2_r}\bigg(|\p_r\psi_r|^2+\frac{1}{r^2}\left(|\p_\theta\psi_r|-A\zeta_r|\psi_r|\right)^2\bigg)\bigg]\ ,
\end{split}
\ee
(using $|z|\geq \Re(z),\Im(z)$).  As the integrand of  ${\mathcal I}$  is homogeneous in  $\psi_l$ and $\psi_r$, 
 the absolute minimum of the integral is given by  $\mathcal{I}=0$,  at  $\psi_l=\psi_r=0$.   
Going back to our Ansatz (\ref{color1}),  we see that such minimum corresponds to constant profile functions  $\rho_l=-1/2g_l$, $\rho_r=-1/2g_r$ and $\eta_l=\eta_r=0$,
 and thus, from Eqs.~(\ref{color1})-(\ref{color3}), one sees that 
\be   
A_{\alpha}^{(l)} = - \frac{ i}{2 g_l}  \de_\alpha (UTU^{\dagger})  (UTU^{\dagger})\ , \qquad    
A_{\alpha}^{(r)} = - \frac{ i}{2 g_r}  \de_\alpha (UTU^{\dagger})  (UTU^{\dagger})\ ,
\ee
is  a pure $SU(N)_{l+r}$ gauge form,
\be    0=  \de_{\alpha} Q^{(B=0)}   \to   D_{\alpha} (UTU^{\dagger})=0\ . 
\ee
In addition,
\be    \mathcal{S}_{eff} =   {\cal I} \,  \int \, dz \,dt  \,  \Tr(\p_\alpha T^{(U)})^2   =0\ ,
\ee
as it should.
Our interest is in excitations of the above static vortex configuration, with $T= 2 \pi \xi$.  Therefore we shall assume 
$\psi_l \ne0$, $\psi_r \ne 0$ in what follows.

Inserting the partial wave expansion (\ref{Ansatz}),  $\mathcal{I}$  becomes  the sum of different angular momentum excitations:
\be\begin{split}
\mathcal{I}&=2\pi\int dr\, r\sum_m\Bigg[\frac{1}{4g_l^2}\left((\p_r\phi_l^m)^2+\left(\frac{m-A\zeta_l}{r}\right)^2(\phi_l^m)^2\right)+\\&+\frac{1}{4g_r^2}\left((\p_r\phi_r^m)^2+\left(\frac{m-1+A\zeta_r}{r}\right)^2(\phi_r^m)^2\right)+\\&+\frac{Q_1^2+Q_2^2}{8}((\phi_l^m)^2+(\phi_r^m)^2)-\frac{Q_1Q_2}{2}\phi_l^m\phi_r^m\Bigg]\ . \label{generalm}
\end{split}\ee
For large $r$  this becomes
\be
\label{div}
\mathcal{I}\sim \frac{2\pi}{4g'^2\zeta_l(1-\zeta_l)}\int\,dr\,r\sum_m\left((\p_r\phi_l^m)^2+\left(\frac{m-\zeta_l}{r}\right)^2(\phi_l^m)^2\right)\ .
\ee
The contribution to  the integral from a single mode 
 \be  \phi_l^m\sim   \phi_r^m\sim   r^{s}\ ,  
\qquad  s=|m-\zeta_l|\ ,    \label{plusminus}\ee
 is therefore
\be \mathcal{I}\sim R^{ 2  |m-\zeta_l|  }\ ,  \qquad      \label{general} \ee  
where $R$ is an infrared cutoff. As the contributions to the energy of the various modes are decoupled, the excitation of minimum energy has  $m$ corresponding to the minimum value of
$s$.
 
%Which wave is the dominant one (with least energy) depends therefore on the relative values of the gauge coupling constants. 
For $g_r<g_l$ we have   \[   1> \zeta_l= \frac{g_l^{2}}{g_l^2+ g_r^2}   >1/2\ ,\]  and thus   the   mode with least energy is the one with $m=1$ for which the divergence is
\be  {\cal I}\sim   R^{ 2 (1- \zeta_l) }=  R^{ 2  \zeta_r }\ .  \label{concluding}\ee
The  solution is given by $\psi_l=\phi_l\, e^{i\theta}$,   $\psi_r=\phi_r$ with $\phi_l$ and $\phi_r$   (dropping the index $m=1$)  satisfying the system
\be
\begin{cases}
\begin{split}\Delta_r\phi_l-\left(\frac{1-\zeta_l A}{r}\right)^2\phi_l-g_l^2\left(\frac{Q_1^2+Q_2^2}{2}\phi_l-Q_1 Q_2\phi_r\right)=0\end{split} \ ,  \\
\begin{split}\Delta_r\phi_r-\left(\frac{\zeta_r A}{r}\right)^2\phi_r-g_r^2\left(\frac{Q_1^2+Q_2^2}{2}\phi_r-Q_1 Q_2\phi_l\right)=0\ .\end{split}
\end{cases}     \label{Nsolns}
\ee

In concluding that the energy diverges as  (\ref{general}), we have assumed  implicitly (\ref{plusminus}), that is that the $\phi_{l, r}^m$  has the growing power law behavior. This  does not necessarily follow from the general  result (\ref{divphil})-(\ref{divphir}).   This can however be shown as follows.
Multiply equation $\eqref{e1}$ by $\phi_l^m$ and  $\eqref{e2}$ by $\phi_r^m$, divide the first equation by $g_l^2$ and the second by $g_r^2$ and  sum the two equations.  Integrating this over the $xy$ plane, one gets
\be  \begin{split}
&\frac{1}{g_l^2}\phi_l^m\, r\,\p_r\phi_l^m\bigg{|}_0^\infty+\,\frac{1}{g_r^2}\phi_r^m\, r\,\p_r\phi_r^m\bigg{|}_0^\infty=\int_0^\infty\,dr\,r\Bigg(\frac{1}{g_l^2}\big[(\p_r\phi_l^m)^2+\left(\frac{m-A\zeta_l}{r}\right)^2  (\phi_l^m)^2\big]+\\&\frac{1}{g_r^2}\big[(\p_r\phi_r^m)^2+\left(\frac{m-1+A\zeta_r}{r}\right)^2 (\phi_r^m)^2\big]+\frac{Q_1^2+Q_2^2}{2}((\phi_l^m)^2+(\phi_r^m)^2)-2Q_1Q_2\phi_l^m\phi_r^m\Bigg)  \\
&=   \frac{2 }{\pi} \,{\cal I}\ .  \label{good}  
\end{split} \ee
Namely, we have reproduced the well known result that any action quadratic in the fields with the standard kinetic term, computed at its minimum, is a total derivative and thus  is determined by the boundary values of the fields only.  
Since the contributions from the lower limit vanish on the left hand side  ($\phi_l^m, \phi_r^m$ must be regular)   and as the right hand side is positive definite,  $\phi_l^m \sim  \phi_r^m$ cannot 
vanish at $r=\infty$.

\begin{figure}[h!]
\begin{center}
\includegraphics[width=3.3in]{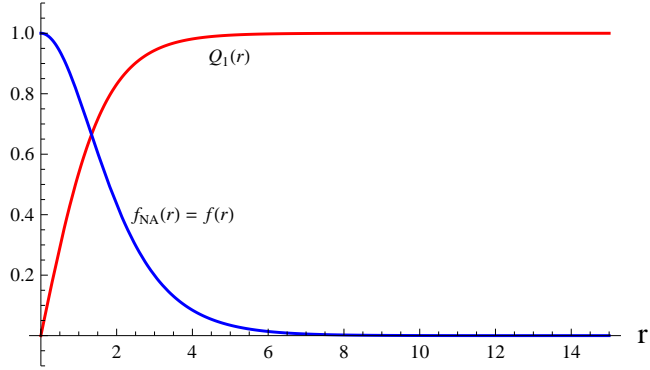}
\caption{Vortex profile functions for  $2 N g_0^2={g'}^2 $.  We take the parameters normalized to $\xi=g'1$. }
\label{vortexprofile}
\end{center}
\end{figure}
\begin{figure}[h!]
\begin{center}
\includegraphics[width=3in]{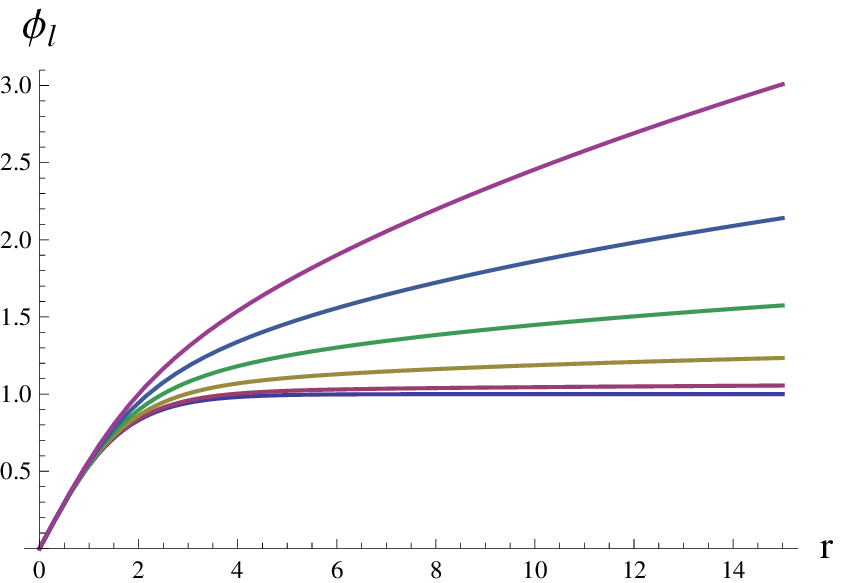}
\includegraphics[width=3in]{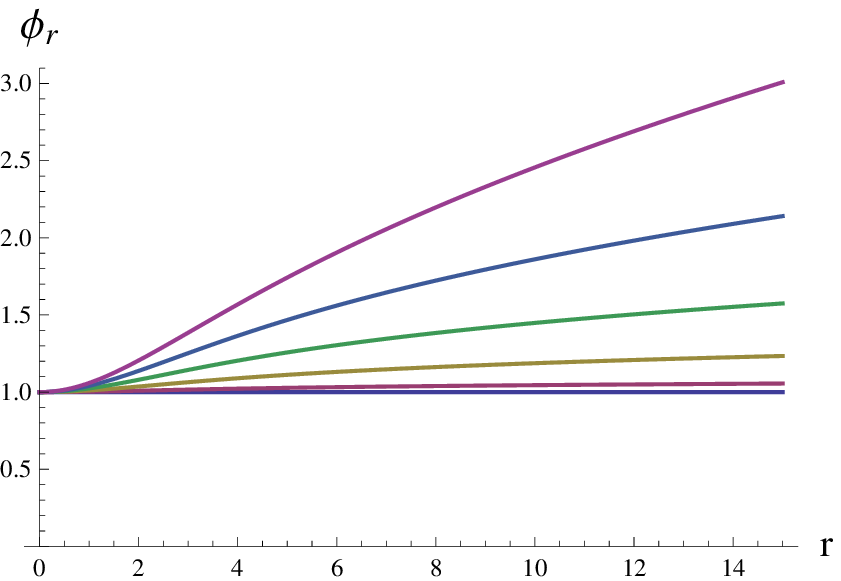}
\caption{The numerical solutions for  $\phi_l$ and $\phi_r$ following from Eq.~(\ref{exactsolphi}),  for various values of $(g_l, g_r)=  (\cos{\theta},\sin{\theta})$ with $\theta = \pi/2 \cdot (0; 0.1; 0.2; 0.3; 0.4; 0.5)$  (from the bottom to the top).  The normalization is fixed  by choosing 
 $\gamma$ such that  $ {\phi_r} (0)=1$. }
\label{phiplot}
\end{center}
\end{figure}
\begin{figure}[h!]
\begin{center}
\includegraphics[width=3.3in]{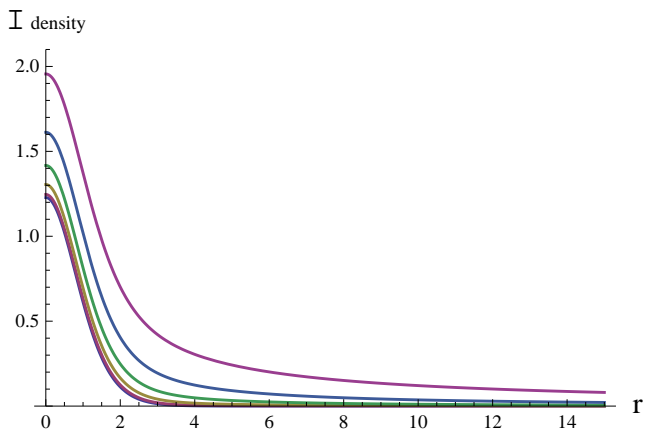}
\caption{The integrand of  ${\mathcal I}$ (\ref{generalm}) as a function of $r$ is given here  for the same set of values of  the coupling constants as in Figure \ref{phiplot}. 
}
\label{energy}
\end{center}
\end{figure}
\begin{figure}[h!]
\begin{center}
\includegraphics[width=3.3in]{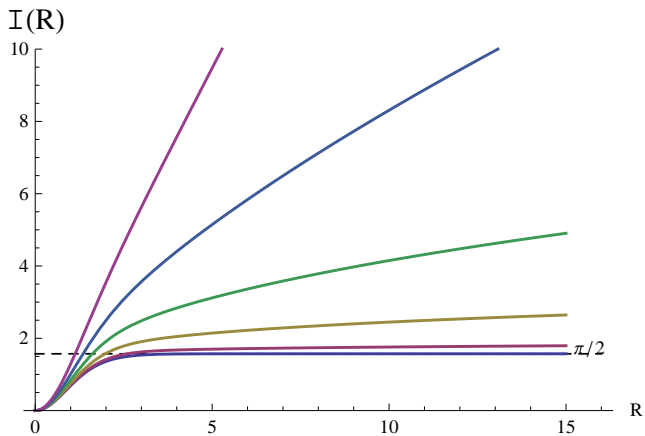}
\caption{${\mathcal I}(R)$ as defined in (\ref{Icutoff}). 
}
\label{IR}
\end{center}
\end{figure}

 The minimum excitation energy ($m=1$ wave for $g_r<g_l$)  corresponds to the particular choice 
\begin{eqnarray}
 \chi_r(z)={\rm const}.\ ,   \qquad {\bar \chi}_l({\bar z})=0\ ,
\end{eqnarray}
in Eq.~(\ref{exactsolution}).
The exact solution of (\ref{Nsolns}) is then
\begin{eqnarray}
\phi_l =  \left(\frac{Q_1}{Q_2}\right)^{\zeta_l} (\gamma r)^{\zeta_r} \ , \qquad \phi_r = \left(\frac{Q_2}{Q_1}\right)^{\zeta_r} (\gamma r)^{\zeta_r} \ .
\label{exactsolphi}
\end{eqnarray}
where $\gamma$ is a constant.
Note that in the limit $g_r \to 0$ ($\zeta_r \to 0$,\, $\zeta_l \to 1$)   this minimum-energy solution smoothly approaches  the known profile function  for the standard, ungauged nonAbelian vortex, brought to the regular gauge form, Eq.~(\ref{ungauged}).

We present below some numerical solutions for the vortex profile functions and the corresponding zero modes. We consider for simplicity the case (\ref{simplifycouplings})  for which the profile functions simplify to (\ref{simplifyvortex}).
For this set of solutions, the vortex profile functions are given in Figure \ref{vortexprofile}.
Also,  $\phi_{l}$ and $\phi_{r}$ given by  (\ref{exactsolphi})    are plotted for several different values of $g_l$ and $g_r$ for constant $g'$,
in Figure \ref{phiplot}.      In Figure \ref{energy} the   planar density in ${\cal I}$  is plotted. 
% in (\ref{generalm}).
The integral ${\cal I}$ can be expressed  by using (\ref{good}),   up to a certain infrared cutoff $R$, as
\beq
\label{Icutoff}
 {\cal I}(R) = \frac{\pi}{2 g_l^2}\phi_l\, r\,\p_r\phi_l\bigg{|}_0^R+\,\frac{\pi}{2 g_r^2}\phi_r\, r\,\p_r\phi_r\bigg{|}_0^R \ .
\eeq
This is  plotted  in Figure \ref{IR}  as a function of $R$ for the particular solution (\ref{exactsolphi}). For $R \gg 1/g' \sqrt{\xi}$ it can be approximated, up to exponential vanishing terms, by
\beq
\label{Icutoffdue}
 {\cal I}(R) \simeq \frac{\pi}{2 g_l^2}  \left(g' \sqrt{\xi} R\right)^{2 \zeta_r} + {\cal O}\left( e^{-g' \sqrt{\xi} R }\right) \ .
\eeq

For $g_r>g_l$ instead,   $\zeta_l<1/2$, so  the excitation energy is minimum for the $m=0$ mode, for which   $\phi_l\sim$ const. and $\phi_r\sim r$ near   $r\to 0$, and   $\phi_l\sim \phi_r\sim r^{\zeta_l}$ at  $r\to\infty$. One has also  in this case 
\be  {\cal I}\sim   R^{ 2  \zeta_r } \ ,  \label{thesame}\ee
which is the same as in (\ref{concluding}). 
The solution is $\psi_l=\phi_l$ and $\psi_r=\phi_r e^{-i\theta}$ with $\phi_l$ and $\phi_r$ satisfying
\be
\begin{cases}
\begin{split}\Delta_r\phi_l-\left(\frac{\zeta_l A}{r}\right)^2\phi_l-g_l^2\left(\frac{Q_1^2+Q_2^2}{2}\phi_l-Q_1 Q_2\phi_r\right)=0\end{split}\ , \\
\begin{split}\Delta_r\phi_r-\left(\frac{1-\zeta_r A}{r}\right)^2\phi_r-g_r^2\left(\frac{Q_1^2+Q_2^2}{2}\phi_r-Q_1 Q_2\phi_l\right)=0\ .\end{split}
\end{cases}
\ee
These are the same as equations (\ref{Nsolns}) in which $\phi_l$ and $\phi_r$ are interchanged.

\subsection{Infrared cutoff}

We have introduced above an infrared cutoff $R$ in the $(x,y)$ plane, to regularize the integral ${\mathcal I}$ in (\ref{Icutoff}).
The divergences arise due to the presence of the massless $SU(N)_{l+r}$ gauge fields in the bulk. There are certain limitations to the parameter space where the effective action in (\ref{Action})-(\ref{eff}) can be applied. First of all we neglected in (\ref{Action}) the terms coming from $F^{l/r}_{\alpha\beta}$ and we have to check when this approximation is reliable. The extra term that we are neglecting in the action is 
\begin{eqnarray}
 \Tr {F^{l/r}_{\alpha\beta}}^2 = \frac{1}{8 g^2_{l/r} } \left( |\psi_{l/r}|^2 - 1 \right)^2  \Tr \left( \p_\alpha T'\p_\alpha T' \p_\beta T'  \p_\beta T' -  \p_\beta T' \p_\alpha T' \p_\beta T' \p_\alpha T'\right) \;.
\end{eqnarray}
This is a higher derivative correction to the effective action, it is thus proportional to $1/\lambda^4$ where $\lambda$ is the typical wavelength of the fluctuation of the $2D$ sigma model we are considering. In order for this terms to be negligible with respect to the two-derivative term the effective action (\ref{effectiveAc}) we need (neglecting some irrelevant coefficients)
\be 
\frac{(g^2_{l}+g^2_{r})  R^2 \left(\left(g' \sqrt{\xi} R\right)^{2 \zeta_r}-\zeta_r-1\right) }{ g^2_{l}g^2_{r} \lambda^4 } \ \ll \ 
\frac{ \left(g' \sqrt{\xi} R\right)^{2 \zeta_r}}{ g_l^2 \lambda^2}  \ .
\ee
This is thus a low-derivative condition, but related to the choice of the cutoff $R$:
\be  \frac{R}{\sqrt{\zeta_r}} \ \ll \ 
\lambda \ .
\ee

Another restriction comes from the fact that the $SU(N)$  unbroken gauge theory is asymptotically free (unless, e.g.,  one introduces many fermions) and becomes strongly coupled at large distances.
Our lowest-order calculations are thus valid only for wavelengths  
much less than the confinement length,  $1/\Lambda_{4D}.$  
At the same time,  by definition of the vortex effective action,  one is calculating  the fluctuations at length scales much larger than the vortex size, $1/ g' \sqrt{\xi}$.
To summarize, one must assume  
\be
\frac{1}{g_r \sqrt{\xi}}   \ \ll \  \frac{R}{\sqrt{\zeta_r}} \ \ll \  \lambda\  \ll \ \frac{1}{\Lambda_{4D}}
\ee
for the validity of our analysis.

\section{Origin of the non-integer  power divergences  \label{sec:origin}}
In this section it will be shown 
that the non-integer power divergences just found are intimately related to the geometric obstruction discussed in Section~\ref{sec:topological}. Our theory has vortex solutions as a consequence of the bulk symmetry breaking $G\to H$,   $\pi_1(G/H) \ne 1$.  When $H$ is nonAbelian, a given vortex configuration further breaks $H$ into 
${\widetilde H} \subset H$: the action of the unbroken generators  in $H/{\widetilde H}$  generates the internal zeromodes. 

We recall also that  the presence of the vortex makes the embedding $H\subset G$ position-dependent in the regular gauge. As a consequence, in a covariantly constant basis some of the generators become multivalued. Multivalued symmetries give rise to Aharonov-Bohm scattering of gauge bosons at large distances from the vortex. The corresponding zero modes lead to nontrivial effects, like cosmic string color superconductivity. They also explain the precise nature of the non-integr, powerlike divergences encountered.

Let us summarize these features in a model independent fashion following the reasoning of Alford et al. \cite{Alford:1990mk}, \cite{Alford:1990ur}. Consider a generic Lagrangian
\be\label{lagr}
\mathcal{L}=-\frac{1}{2}\Tr(F_{\mu\nu}F^{\mu\nu})+|D_\mu Q|^2-V(Q)\ ,
\ee
which describes a gauge theory with gauge group $G$ coupled to a scalar field $Q$, with $D_\mu=\p_\mu-ig A_\mu$ and $A_\mu=A_\mu^a T^a$, where $T^a$ are the generators of $G$ in the $Q$ representation, normalized as $\Tr(T^a T^b)=\delta_{ab}/2$. The scalar potential is chosen so that $Q$ condenses, breaking the full gauge group $G$ to a subgroup $H$. For a nontrivial  $\pi_1(G/H)$ vortex configurations exist. The elementary vortex is oriented in some fixed direction in the Lie algebra that we denote as $S$. One may take the following Ansatz
\be
Q=u(\theta)  Q_0(r)\ ,\qquad
A_r=0\ ,\qquad
A_\theta=a_\theta(r) S\ ,\qquad
A_\alpha=0\ ,
\ee
where $\alpha=z,t$.
$Q_0$ approaches a uniform vacuum configuration at spatial infinity and $u(\theta)$ describes the asymptotic winding of the scalar condensate,
\be
u(\theta)=e^{i\theta S}\ .
\ee
The Ansatz for the gauge fields is determined by the requirement of finiteness of the  energy. In particular, a necessary condition is that $|D_iQ|\to 0$ as $r\to\infty$, which implies $A_r\to 0$ and $A_\theta\to (1/g)\,S$. Therefore, the boundary conditions on the profile function $a_\theta$ are $a_\theta\to1/g$ as $r\to\infty$ and $a_\theta\to 0$ as $r\to 0$. Since the vacuum of the theory leaves an unbroken symmetry group, which however can act nontrivially on our Ansatz, the vortex carries orientational moduli, describing its embedding in the Lie algebra. Through a global gauge transformation $U$ of $H$, one obtains a vortex with a generic orientation,
\be
Q=U u(\theta)  Q_0(r)\ ,\qquad
A_r=0\ ,\qquad
A_\theta=a_\theta(r) S^{(U)}\ ,\qquad
A_\alpha=0\ ,
\ee
where $S^{(U)}=USU^\dagger$. 
%For different choices of $U$, one has physically distinct configurations.
%The subset of global gauge transformations represents real symmetries of the system, not just a redundancy of description.
In the spirit of the moduli space approximation, $U$ is taken to depend on the string worldsheet coordinates $t$ and $z$. Then, 
\be
Q=U(t,z) u(\theta)  Q_0(r)\ ,\qquad
A_r=0\ ,\qquad
A_\theta=a_\theta(r) S^{(U)}(t,z) \ ,\qquad
A_\alpha\ne 0\ .
\ee
As soon as $U$ is taken to fluctuate along the string, nontrivial $A_\alpha$ fields are induced, whose precise form is dictated by the equations of motion. 
Their behavior at spatial infinity, however, is fixed by the requirement $|D_\alpha Q|\to 0$, which implies that    $A_{\alpha}$  approaches (the $U$ transform of)
the gauge fields belonging to $H \subset G $  left unbroken by the vacuum configuration  $u(\theta)  Q_0(\infty)$.

We may now gauge transform by $U^\dagger(t,z)$, going to what we shall call  the {\it static gauge}. This gauge choice makes computation somewhat  simpler. The new Ansatz is
\be
Q= u(\theta)  Q_0(r)\ ,\qquad
A_r=0\ ,\qquad
A_\theta=a_\theta(r) S \ ,\qquad
A_\alpha\ne 0\ .
\ee 
One obtains for $F_{i\alpha}$
\be
F_{i\alpha}=\p_iA_\alpha-iga_i[S,A_\alpha]\ ,
\ee
where $A_i=a_i(r)S$. Passing to the static gauge allowed us to eliminate the $\p_\alpha A_i$ piece. 
The equations of motion following from the Lagrangian density $\eqref{lagr}$ are
\be\label{eomt}
D^\mu D_\mu Q=-\frac{\p V}{\p Q^\dagger}\ , \qquad 
\mathcal{D}_i F_{i\alpha}=igT^a Q^\dagger T^a D_\alpha Q+\text{h.c.}\ .
\ee
We focus now on the second of equations $\eqref{eomt}$, which is the nonAbelian Gauss law. Substituting it into our Ansatz, the left hand side becomes
\beq
\mathcal{D}_i F_{i\alpha}&=&\Delta  A_\alpha-2ig\frac{a_\theta}{r^2} [S,\p_\theta A_\alpha]-g^2\frac{a_\theta^2}{r^2} [S,[S,A_\alpha]] \nn
&=&\Delta_r A_\alpha +\frac{1}{r^2}\left(\p_\theta -ig a_\theta [S,\cdot]\right)^2A_\alpha\ .    \label{nonzero}\eeq
 This rewriting also follows from the fact that in static gauge $F_{i\alpha}=\mathcal{D}_i A_\alpha$ and 
 \be
 \mathcal{D}_i F_{i\alpha}=\mathcal{D}_i\mathcal{D}_i A_\alpha=\Delta_r A_\alpha+\frac{1}{r^2}\mathcal{D}_\theta^2 A_\alpha\ ,
 \ee
where the covariant derivatives in $r$ reduce to the ordinary derivatives since $A_{ r}=0$. 
At large $r$ the right hand side of Gauss's equation vanishes as  the covariant derivative of the scalar field approaches zero. 
 One obtains  thus a Laplace-type equation
\be\label{lapltyp}
\Delta_r A_\alpha\simeq -\frac{1}{r^2}(\p_\theta-i[S,\cdot])^2 A_\alpha\ .
\ee
One may expand $A_\alpha$ in eigenstates of the operator $\Delta_\theta=\p_\theta-i[S,\cdot]$, which at spatial infinity commutes with $\Delta_r$,
\be\label{eig}  
\Delta_\theta \, \psi_\alpha^s=is\psi_\alpha^s\ ,
\ee
with  $s$ being a real contant.
The eigenfunction in $\eqref{eig}$ can be written as 
\be\label{ptsol}
\psi_\alpha^s=e^{is\theta}u(\theta) A_\alpha(r,0) u(\theta)^\dagger\ ,
\ee
using the fact that
\be
(\p_\theta-i[S,\cdot])e^{i\theta S} M e^{-i\theta S}=0\ ,
\ee 
for any $\theta$ independent matrix $M$.\footnote {This is simply the condition of parallel transport of a constant.} If a parallel transport by $u(\theta)$ describes the embedding of $H$ inside $G$, we see that if $A_\alpha$ at $\theta=0$ is oriented along some generator $S^a$, at $\theta\neq 0$ it is oriented along the rotated generator $S^a(\theta)$.

One can now see  the consequences of the requirement of single valuedness of $\psi_\alpha^s$. If $S^a$ is covariantly constant and belongs to the globally defined subgroup $\widetilde H\subset H$, then
\be
u(2\pi)S^au(2\pi)^\dagger=S^a\ ,
\ee
and the  single-valuedness condition for  $\psi_\alpha^s$ implies that $s\in \mathbb{Z}$. If, on the contrary, $S^a$ is not globally defined,
\be
u(2\pi)S^au(2\pi)^\dagger=e^{2\pi i \xi^a}S^a\ ,   \label{corresp}
\ee
with some  non integer constant $\xi^a$.  The single valuedness of $\psi_\alpha^s$ then implies that  
\be  s\in \mathbb{Z}-\xi^a\ .\ee
 Using $\eqref{ptsol}$ in $\eqref{lapltyp}$, one finds the asymptotic behavior, 
\be
\Delta_r \psi_\alpha^s\simeq \frac{s^2}{r^2} \psi_\alpha^s\ \qquad  \Longrightarrow  \quad \  \psi_\alpha^s
 \propto r^{\pm s}\ . 
\ee
The energy grows like $R^{2|s|}$. Allowing for a 
nonvanishing angular momenta  $m \in  {\mathbb Z}$ in   (\ref{nonzero}), (\ref{lapltyp}),  one gets in general 
\be     A_{\alpha} \sim  {\cal O}(R^{|s-m|} )\ , \qquad  {\rm Energy} \sim   {\cal O}(R^{2|s-m|})\ .\label{predict}
\ee

\vskip 4mm

Such a general argument, however,  carries us only this far.  In particular, the  determination of the generators  along which the fields $A_\alpha$ are excited can only be done by studying the nonAbelian Gauss equation {\it   for all $r$} and requiring that the color matrix structures of the left and right hand sides match precisely,  as has been done in Sections~\ref{sec:excitation}  and \ref{sec:energy}.  Related to this, the non-integr power $s$ 
cannot be determined by a general argument based on the asymptotic behaviors of the fields alone.   Both of these features carry information about the vortex configuration at finite $r$.

In our specific $U_0(1)\times SU_l(N)\times SU_r(N)$  theory,  
it was found   
%(Section~\ref{sec:excitation}  and Section~\ref{sec:energy})   
that  the gauge fields excited by the vortex modulation  lie in the directions
\be\label{orans}
A_\alpha^{(l)}=\rho_lW_\alpha+\eta_lV_\alpha\ ,\qquad  
A_\alpha^{(r)}=\rho_rW_\alpha+\eta_rV_\alpha\ .
\ee
with
\be \begin{split} &W_\alpha \equiv   i \, \p_\alpha T^{(U)} T^{(U)} \ , \qquad  V_{\alpha}\equiv \p_\alpha T^{(U)} \ ,    \\
&T^{(U)}  =  U T U^{\dagger}\ ,  \qquad\quad\quad   T=  \left(\begin{array}{cc}1 &  \\ & -{\bf 1}_{N-1}\end{array}\right)\  .    \end{split}\label{color3bis}
\ee
Knowing this, and knowing how these gauge fields are parallel-transported around the vortex   (see (\ref{g1})), 
\be   A_\alpha^{(l)}   \to    \mathpzc{u}_l(\theta) A_\alpha^{(l)}  \mathpzc{u}_l(\theta)^{\dagger}\ , \qquad       \mathpzc{u}_l(\theta) \equiv  e^{i\zeta_l\frac{\theta}{NC_N}T^{(U)}_{N^2-1}}\ ,
\ee
(similarly for $A_\alpha^{(r)} $),   one finds that 
upon encircling the vortex
\be   \label{monodromy}
\left(\begin{array}{c}
W_\alpha \\
V_{\alpha}\\
\end{array}\right)
\to \left(\begin{array}{cc}
\cos(2\pi\zeta_l)&\sin(2\pi\zeta_l)\\
-\sin(2\pi\zeta_l)&\cos(2\pi\zeta_l)\\
\end{array}\right) \left(\begin{array}{c}
W_{\alpha}\\
V_{\alpha} \\
\end{array}\right)\ ,
\ee
where the commutation relations 
\be
[W_\alpha,T^{(U)}]=2i V_\alpha\ ,\qquad
[V_\alpha,T^{(U)}]=-2i W_\alpha\ ,
\ee
have been used. 
The eigenvalues of the monodromy matrix  (\ref{monodromy})  are thus   $e^{- 2\pi i\zeta_l}$ and $e^{2\pi i\zeta_l}$, with respective   eigenvectors  $v=\p_{\alpha}T^{(U)} T^{(U)}+\p_{\alpha}T^{(U)}$ and its Hermitian conjugate $v^\dagger$. Thus in our model  $\xi^a$ in equation $\eqref{to}$ or $\eqref{corresp}$ is given concretely  by 
\be \xi^a = \pm \zeta_l= \pm \frac{g_l^2}{g_l^2+ g_r^2}\ . \ee
Our results in Section~\ref{sec:energy}   indeed agree with the general expectations (\ref{predict}) with this specific 
index. 

These discussions clearly illustrate the connection between the global features of the system discussed in Section~\ref{sec:topological}  and the dynamical properties of the vortex excitation  explored in Section~\ref{sec:energy}.  
In particular,  let us note that for the standard nonAbelian vortex ($g_r=0$ or $g_l=0$)    $\xi^a$ reduces to an integer. Thus no geometric obstruction, no nonAbelian AB effects, no non-integr-power growth of vortex excitation energy, etc.,  occur.  
 The $2D$  $\mathbbm{CP}^{N-1}$ dynamics becomes fully operative, and at the same time the massless, free 
$SU_r(N)$   (or     $SU_l(N)$) Yang-Mills system in $4D$   simply decouples from the vortex system.

\section{Discussion \label{sec:discussion}}

In this paper we discussed in some detail the topological and dynamical features of an interacting $4D$-$2D$ coupled quantum field theory, in the context of the underlying  $U_0(1)\times SU(N)\times SU(N)$ gauge theory in an $SU(N)_{l+r}$ symmetric  ("color-flavor" locked) vacuum.  Of course,  such a system can be considered {\it ab initio}, just by starting with a given $2D$ system possessing some global symmetry, and making it local, by introducing $4D$ gauge fields  coupled to it \cite{Hull, Gaiotto}, i.e., embedding the $2D$ system in $4D$ spacetime. 
Another venue, which we have opted to follow here, is to study such a $4D$-$2D$ coupled quantum field theory  which emerges as a low-energy effective description of the vortex sector of the underlying $4D$ system.  Here the $2D$ variables are the fluctuations of the vortex  collective coordinates, whereas the $4D$ degrees of freedom are part of the original Yang-Mills fields, the coupling among them being  fixed by the local gauge symmetries of the underlying  theory.    

 This is a new type of interacting quantum field theory, in which field variables living in different dimensions interact nontrivially. 
 Our way of approaching the problem gives an existence proof that such a theory can be consistently defined, as the underlying $4D$ system is 
 a consistent local gauge theory.

In a standard nonAbelian vortex, the modulation of the vortex internal orientation is described by  a $2D$  sigma model ($\mathbbm{CP}^{N-1}$ in the case of $U_0(1)\times SU(N)$  theory) on the vortex worldsheet. The $\mathbbm{CP}^{N-1}$ model in two dimensions is asymptotically free and becomes strongly coupled at low energies  $ \Lambda \ll \sqrt{\xi}$.  The renormalization group flow  of two dimensional model  ``carries on"  the evolution  of the $4D$  gauge coupling  which was frozen at the vortex mass scale $\sqrt{\xi}$, into the infrared,   showing a remarkable realization of  $2D$-$4D$  duality \cite{Dorey:1998yh,Hanany:2003hp,Gorsky:2004ad}.

%The low energy regime  $\ll \sqrt{\xi}$, is also the regime in which the effective action for the $B$ modes was derived. 
When the orientational mode dynamics becomes strongly coupled, the orientation $\{ B_i \}$ of the vortex in the Lie algebra fluctuates  wildly,  and the AB effects are washed away. This is consistent with the fact that there is no spontaneous symmetry breaking and no Goldstone bosons in 2D: the spectrum has a mass gap.
%there are no Goldstone bosons and the orientation of the nonabelian vortex is evolving along the string with nontrivial dynamics. 

As soon as  $g_r$ is turned on, however, the physics changes immediately, and drastically. 
The presence of massless fields in the bulk makes the coefficient in front of the effective $\mathbbm{CP}^{N-1}$  action divergent and the vortex fluctuations become
nonnormalizable.  They become infinitely costly to excite.   At the same time,   the AB effects and all sorts of topological and nonlocal  effects emerge. 

 \vskip 4mm
 
 One wonders however how such a discontinuous change of physics is possible at $g_r \to 0$.  As in the phenomena of phase transitions,  
 physics cannot really exhibit such a qualitative change  at $g_r \to 0$, if the spacetime is  finite \footnote{Here the relevant dimension is the extension of the transverse $(x,y)$  space $R$.}. The expression (\ref{concluding}) indeed shows that the limits $g_r \to 0$ and $R \to \infty$  do not commute, pointing clearly to the origin of the %apparent 
discontinuity.  Stated differently, 
 it is because one is accustomed to thinking about infinitely extended spacetime in relativistic-field-theory applications  that physics looks discontinuous at $g_r=0$.

 \vskip 4mm

Let us look into the nature of this discontinuity more carefully.  At large $r$  the divergent effects such as 
Eq.~(\ref{plusminus})-Eq.~(\ref{concluding})  come from the dominant,  massless $SU_{l+r}(N)$ gauge field excitations  (the massive fields in the bulk Higgs mechanism give  only exponentially suppressed contributions).    They {\it are} nonetheless the result of the  interactions with the  $2D$  vortex collective coordinates, the fluctuations of $\{ B_i \}$, and in fact,  the non-integr power  and the precise direction in color space in which the massless gauge fields are excited, both reflect the details of the vortex configurations near  the vortex core.

What happens in the $g_r \to 0$  limit is that most of these divergent effects smoothly  transit into the continuum spectrum of the free $SU_r(N)$ Yang-Mills system, now decoupled from the vortex.  The fact that they appear  in the coefficient in front of the vortex effective action
 \beq
S_{eff} &=&
\int \,dt  d^{3}x\,  [\Tr(F_{i\alpha}^{(l)}F^{(l)\,\,\alpha}_{\,\,\,\,\,\,i})+\Tr(F_{i\alpha}^{(r)}F^{(r)\,\,\alpha}_{\,\,\,\,\,\,i})+\Tr|D_\alpha Q|^2]\, \nn &=&  
  {\cal I} \,  \int \, dz \,dt  \,  \Tr(\p_\alpha T^{(U)})^2  
\label{ActionBis}\eeq
 does not necessarily mean that they are related to the vortex physics.

 Summarizing, if one first takes the IR cutoff $R\rightarrow\infty$, the screening of the massive gauge fields implies that the massless gauge fields dominate, leading to the power law scalings.
On the other hand, if one  takes the limit $g_r\rightarrow 0$ first, 
for the vortex effective action of {\it  minimum energy cost}, i.e., the contribution of  the partial $m=1$ wave,    one finds from (\ref{Icutoffdue}) that 
 \be  \lim_{g_r \to 0}{{\cal I}(R)} = \frac{\pi}{2 g_l^2}  + {\cal O}\left( e^{-g_l \sqrt{\xi} R }\right) \   \label{concludingBis} \ .\ee
Taking then the $R \to \infty$ limit we recover the physics of  the standard nonAbelian vortices with a finite $\mathbbm{CP}^{N-1}$ coupling 
 \be   {\cal I} = \frac{\pi}{2  g_l^2}   \ .
 \ee 
Taking the limits in this order, the only  massless degrees of freedom that remains are the $\mathbbm{CP}^{N-1}$ orientational modes in the vortex worldsheet.

 \vskip 4mm

The renormalization of the $\mathbbm{CP}^{N-1}$ model, for the ungauged nonAbelian vortex, is given by
\be
8 \pi {\cal I}(\mu) = \log{\left( \frac{\mu}{\Lambda_{2D}} \right)}
\ee
namely ${\cal I}$ is {\it decreasing} logarithmically as the RG scale $\mu$ decreases.
For the gauged nonAbelian vortex, we find another effect: ${\cal I}$ is {\it increasing} as the cutoff scale $1/R$ gets smaller, this time as a power law (\ref{Icutoffdue}), due to the interaction with the massless unbroken $SU(N)$.  
We may thus conjecture the $\mathbbm{CP}^{N-1}$ model coupled to the $4D$  massless gauge fields will remain {\it frozen} in the Higgs phase, at least as long as the massless $SU(N)$ field in $4D$ remains in the Coulomb phase.  
Note that  Coleman's theorem on the absence of the Nambu-Goldstone modes in two-dimensional spacetime would not apply here due to the interaction of the $4D$  massless gauge fields.  This would mean that the global and topological effects related to the vortex orientation $\{ B_i \}$ would not be washed away by the $2D$ quantum effects.

To compute the effective action in this paper, we have integrated out all gauge fields, including the massless gauge fields in the bulk. 
This is formally not quite consistent from the renormalization group point of view.  
One should integrate away the massive gauge fields first, and obtain the effective action defined at the vortex mass scale $\sqrt{\xi}$, which should play the role of an UV cutoff for the study of further quantum effects.  The  effective action at $\Lambda_{UV}=\sqrt{\xi}$ would consist of two pieces: the full 4D action for the massless gauge fields living in the bulk and the 2D action describing the fluctuations of the internal orientation,  $\{ B_i \}$, which is a sort of Nambu-Goldstone bosons living in the vortex worldsheet. 
Their coupling is dictated by the full $H \subset G$ gauge invariance.   
 By integrating out the coupled massless $4D$ gauge fields and the $2D$ fields $\{ B_i \}$ simultaneously  down to some infrared cutoff scale  $\mu$, 
 and varying it,  one would find an appropriate RG flow towards the infrared. 

Although such a renormalization group program is still to be properly set up and be worked out, we believe that the number of results established and the subtle issues of decoupling / transitions to the $g_r \to 0$ limit clarified here, should provide us with a useful starting point for such a task.  

\section*{Acknowledgments} 

The authors thank Muneto Nitta for useful discussions at an earlier stage of the work and Simone Giacomelli for interesting discussions and for information.  KK thanks M. Nitta and H. Itoyama for warm hospitality 
during several visits to Keio University and Osaka City University. KO thanks H. Itoyama for all the help given to him at the 
Osaka City University.
 JE is supported by NSFC grant 11375201. This research work
  is supported by the INFN special project GAST (``Gauge and String Theories").

\appendix

\section{Excitation energy in the standard nonAbelian vortex ($g_{r}=0$)\label{gr=0case}}

In this Appendix  we consider the special case $g_r=0$ and discuss  how the analysis of Section~\ref{sec:energy}  gets modified. 
By partial wave expansion
\be \label{cii}
\begin{split}
&\psi_l=\sum_m \phi_l^m e^{im\theta}\,;\\
&\psi_r=1\ ,
\end{split}
\ee
one finds
\be\begin{split}
\mathcal{I}&=2\pi\int dr\,r\Bigg(\frac{1}{4g_l^2}\sum_m\left((\p_r\phi_l^m)^2+\left(\frac{m-A}{r}\right)^2 (\phi_l^m)^2\right)+\\&+\frac{Q_1^2+Q_2^2}{8}\left(\sum_m(\phi_l^m)^2+1\right)-\frac{Q_1Q_2}{2}\phi_l^1   \Bigg)\ .
\end{split}  \label{Iforgr0}\ee
It will be shown below that actually only for $m=1$ there exist physical solutions. 
Substituting accordingly  the  Ansatz  $\eqref{cii}$ in  $\eqref{Iforgr0}$, keeping only the $m=1$ term and dropping the index $l$ for the $SU_L(N)$,  i.e., 
\be
\psi_l=\phi^1  \,  e^{i \theta}  \equiv   \sigma^s e^{i\theta}\ , 
\ee
one gets
\be
\mathcal{I}=\int \,d\theta\,dr\, r\bigg[ \frac{1}{4g^2}\bigg((\p_r\sigma^s)^2+\bigg(\frac{f_{NA}\sigma^s}{r}\bigg)^2\bigg)+\frac{Q_1^2+Q_2^2}{8}((\sigma^s)^2+1)-\frac{Q_1Q_2}{2}\sigma^s\bigg]\ .
\ee
This is a familiar expression for $\mathcal{I}$  found earlier\cite{Auzzi:2003fs}-\cite{Gudnason:2010rm}:  
by making use of the  equations for $\{Q_{1,2}, f_{NA}\}$, it can be shown that at its minimum  
\be
\mathcal{I}=\frac{\pi}{2g^2}\ .     
\ee
Thus for the standard nonAbelian vortex, where $SU_R(N)$  (hence the color-flavor diagonal $SU(N)$)  is a global symmetry, 
there are  no divergences  in ${\cal I} $ in front of the ${\mathbbm {CP}}^{N-1}$ action (Eq.~(\ref{effectiveAc})).  The vortex orientational zeromodes fluctuate strongly in the infrared   in the vortex worldsheet.

Consider instead  the contribution of a  single wave $\phi_l^m $,  $m \ne 1$.  By dropping the indices  $l$ for $SU_L(N)$ and $m$ for the angular momentum   ($\phi\equiv   \phi_l^m $)  one has 
\be\label{normintm}
\mathcal{I}=2\pi \int\,dr\,r\left[\frac{1}{4g_l^2}\left((\p_r\phi)^2+\left(\frac{m-A}{r}\right)^2\phi^2\right)+\frac{Q_1^2+Q_2^2}{8}(\phi^2+1)\right]\ .
\ee
The equation of motion for $\phi$  is
\be\label{preqm}
\Delta \phi=\left(\frac{m-A}{r}\right)^2\phi+\frac{Q_1^2+Q_2^2}{2}g_l^2\phi\ .
\ee
Near the origin $r= 0$    ($Q_1\to 0$, $Q_2\to $ const.   and $A\to 0$)   this becomes
\be
\left(\Delta-\frac{m^2}{r^2}\right)\phi\sim 0\ ,\qquad 
\ee
which has solutions $\phi\sim r^{\pm m}$  (or    $\phi \sim {\rm const.},\,  \log r$  for $m=0$).
As   $r\to \infty$    ($Q_1,\,Q_2\to\sqrt{\xi}$ and $A\to 1$),  Eq.~(\ref{preqm}) becomes 
\be
\Delta \phi\sim \left(\frac{m-1}{r}\right)^2\phi+\mu^2\phi \sim\mu^2 \phi\ ,     
\ee
where $\mu \equiv   g_l\sqrt{\xi}$ is  the W-boson mass.
The solution is a combination of  modified Bessel functions of the first and  second kind, i.e.,  
 at large $r$
\be
\phi\sim \frac{c_{1} \, e^{\mu r} + c_{2}\, e^{-\mu  r}}{\sqrt{r}}\ , 
\ee
with some constants  $c_{1,2}$.
To show that the exponentially growing term is necessarily  present for the solution regular at the origin ($\phi \sim r^{|m|}$),   multiply Eq.~(\ref{preqm}) by $\phi$ and integrate over  the $(x,y)$ plane, to get
\be         \phi \, r\, \de_{r}  \phi |_{0}^{\infty}  =    \int dr  r  \,  \left[  \,  (\de_{r} \phi)^{2} +  \left(\frac{m-A}{r}\right)^2\phi^{2}+\frac{Q_1^2+Q_2^2}{2}g_l^2\phi^{2} \, \right]>0 \ .
\label{procedure}\ee
This would lead to a contradiction, unless  $c_{1}\ne 0$. 

Such an exponentially growing behavior for the gauge fields must be regarded as unphysical (not an acceptable quantum state), and we discard them.

It is amusing to see  what makes the  difference in the special case $m=1$. 
The normalization integral $\mathcal{I}$ is
\be
\label{normint1}
\mathcal{I}=2\pi \int\,dr\,r\left[\frac{1}{4g_l^2}\left((\p_r\phi)^2+\left(\frac{1-A}{r}\right)^2\phi^2\right)+\frac{Q_1^2+Q_2^2}{8}(\phi^2+1)-\frac{Q_1Q_2}{2}\phi\right]
\ee
and the equation of motion is 
\be 
\Delta \phi=\left(\frac{1-A}{r}\right)^2\phi+g_l^2 \frac{Q_1^2+Q_2^2}{2}\phi-  g_l^2 Q_1Q_2\ . \label{exacts}
\ee
The crucial difference is the last term in (\ref{normint1}), (\ref{exacts}).
At   $r\to 0$, $\eqref{exacts}$ gives
\be
\Delta \phi\sim \frac{1}{r^2}\phi\ ,
\ee
whose  regular solution behaves as $\phi\sim r$. 
At   $r\to\infty$, instead,
\be
\Delta \phi=\mu^2(\phi-1)\ ,
\ee
which implies
$
\phi\sim 1+c\,\frac{e^{\pm\mu r}}{\sqrt{r}}.
$
From the exact solution of (\ref{exacts})  we know that actually  $\phi$ behaves as 
\be  \phi\sim 1 -  \frac{e^{- \mu r}}{\sqrt{r}}\ .  \ee  
The procedure which has led to (\ref{procedure})  before yields  this time 
\be         \phi \, r\, \de_{r}  \phi |_{0}^{\infty}  =    \int dr  r  \,  \left[  \,  (\de_{r} \phi)^{2} +  \left(\frac{m-A}{r}\right)^2\phi^{2}+g_l^2  \frac{Q_1^2+Q_2^2}{2}\phi^{2} -
g_l^2  Q_{1} Q_{2}   \phi \, \right] \ .
\label{procedureBis}\ee
Even though the LHS vanishes in this case also, the integrand of the right hand side is no longer positive definite, thus  no contradiction arises. 

%Because of the offset by 1, which really comes from the inhomogenous term of $\eqref{preq1}$, our previous reasoning that the exponentially divergent solution must be present is invalid. As $r\to\infty$, only the damped exponential is present, $\phi$ approaches 1 exponentially fast, the energy density vanishes. In fact, $\mathcal{I}$ is perfectly finite and an exact solution for $\phi$ can be found. The conclusion is that only the $m=1$ mode is normalizable, all other modes have exponential blow-up behavior. 

We conclude that in the $g_{r}=0$ theory,  the only physical mode in Eq.~(\ref{normintm}) is the $m=1$ wave.  
This is to be contrasted with the powerlike divergent modes for the general  gauged vortex,    Eqs.~(\ref{concluding}), Eq.~(\ref{thesame}).
Even though these latter solutions are nonnormalizable also, they represent the continuous spectrum of the theory and as such are to be regarded as physical excitation modes.

\section{Gauge fixing for  $\psi_l$ and $\psi_r$}

We have seen in the main text that  the Ansatzes (\ref{color1})-(\ref{color3}),
\be 
A_\alpha^{(l)}=i\rho_l \, W_{\alpha}+\eta_l\, V_{\alpha}\ ,  \label{color1bis}
\qquad  
A_\alpha^{(r)}=i\rho_r \,   W_{\alpha}+\eta_r \,V_\alpha\ , 
\ee
where 
\be
{\begin{split}
&W_\alpha \equiv \p_\alpha T^{(U)} T^{(U)} \ , \qquad  V_{\alpha}\equiv \p_\alpha T^{(U)} \ ,   \\
&   T^{(U)}  =  U T U^{\dagger}\ ,  \qquad\qquad   T=  \left(\begin{array}{cc}1 &  \\ & -{\bf 1}_{N-1}\end{array}\right)\ ,     %\label{color3bis}
\end{split}}
\ee
 solve the Gauss equation, given the vortex configuration
\be  Q^{(B)}=    U(B) \,  \left(\begin{array}{cc}e^{i \theta} \, Q_1(r) & 0 \\0 & Q_2(r)\, {\bf 1}_{N-1}\end{array}\right)   U^{\dagger}(B)\ .   
\ee  
 One wonders whether this is also necessary, i.e., if there are any other choice for the gauge fields which solve it.   
The equations of motion (\ref{ee}) and (\ref{ef}), or the expression for the energy (\ref{ci}),  are clearly invariant under a common phase rotation,
\be  \psi_l \to e^{ i \beta} \psi_l\ , \qquad  \psi_r \to    e^{ i \beta} \psi_r\ .  
\ee
By recalling the definitions    
\be
\psi_l=\sigma_l+2ig_l\eta_l\ , \qquad   (\sigma_l= 1 + 2 g_l \rho_l)\ , 
\ee
\be
\psi_r=\sigma_r+2ig_r\eta_r\ ,  \qquad   (\sigma_r= 1 + 2 g_r \rho_r)\ , 
\ee
one can however show  that the scalar and gauge fields  
\be    \{  e^{ i \beta} \psi_l, e^{ i \beta} \psi_r, Q (\theta) \}
\ee
are   gauge equivalent to  
\be       \{  \psi_l, \psi_r, Q (\theta + \beta) \}\ .  
\ee
The algebra needed is the same as the one involved in the gauge transformation from the singular to regular gauge,  see Eqs.~(\ref{justagauge})-(\ref{gtpsi}). 
In other words, the apparent extra zeromode  corresponds to the space rotation  (the shift of the origin  of $\theta$): it does not represent 
a physical zeromode.

\end{document}